\documentclass{article}

\usepackage{arxiv}

\usepackage[utf8]{inputenc} 
\usepackage[T1]{fontenc}    
\usepackage[colorlinks,citecolor=blue,urlcolor=black, linkcolor=blue,linktocpage=true]{hyperref}       
\usepackage{float}
\usepackage{url}            
\usepackage{booktabs}       
\usepackage{amsfonts}       
\usepackage{nicefrac}       
\usepackage{microtype}      
\usepackage{graphicx}
\usepackage{amssymb,amsthm}
\usepackage{amsmath}
\usepackage{subcaption}
\usepackage[
backend=biber,
style=authortitle,
citestyle = authoryear,
maxbibnames = 8,
maxcitenames = 2,
natbib = true,
uniquelist=false,
uniquename= false,
]{biblatex}
\DeclareNameAlias{sortname}{family-given}
\newtheorem{assumption}{Assumption}

\theoremstyle{remark}

\newtheorem*{remark}{Remark}

\newcommand{\indep}{\perp \!\!\! \perp}

\title{2D  score based estimation of heterogeneous treatment effects}

\author{
  Steven Siwei Ye \\
  Department of Statistics\\
  University of California, Los Angeles\\
  Los Angeles, CA 90095 \\
  \texttt{stevenysw@g.ucla.edu} \\
     \And
 Yanzhen Chen\\
  Department of ISOM\\
  Hong Kong University of Science and Technology\\
  Clear Water Bay, Kowloon Hong Kong \\
  \texttt{imyanzhen@ust.hk} \\
   \And
 Oscar Hernan Madrid Padilla\\
  Department of Statistics\\
  University of California, Los Angeles\\
  Los Angeles, CA 90095 \\
  \texttt{oscar.madrid@stat.ucla.edu} \\
}

\begin{document}
\maketitle

\begin{abstract}
{Statisticians show growing interest in estimating and analyzing heterogeneity in causal effects in observational studies. However, there usually exists a trade-off between accuracy and interpretability for developing a desirable estimator for treatment effects, especially in the case when there are a large number of features in estimation. To make efforts to address the issue, we propose a score-based framework for estimating the Conditional Average Treatment Effect (CATE) function in this paper. The framework integrates two components: (i) leverage the joint use of propensity and prognostic scores in a matching algorithm to obtain a proxy of the heterogeneous treatment effects for each observation, (ii) utilize non-parametric regression trees to construct an estimator for the CATE function conditioning on the two scores. The method naturally stratifies treatment effects into subgroups over a 2d grid whose axis are the propensity and prognostic scores. We conduct benchmark experiments on multiple simulated data and demonstrate clear advantages of the proposed estimator over state of the art methods. We also evaluate empirical performance in real-life settings, using two observational data from a clinical trial and a complex social survey, and interpret policy implications following the numerical results. The R code for implementing the method introduced in the paper is publicly available on one of the author’s Github page (\url{https://github.com/stevenysw/causal_pp}).}
\end{abstract}

\keywords{observational data \and subgroup treatment effects \and regression trees \and matching}

\section{Introduction}
 
The questions that motivate many scientific studies in disciplines such as economics, epidemiology, medicine, and political science, are not associational but causal in nature. In the study of causal inference, many researchers are interested in inferring average treatment effects, which provide a good sense of whether treatment is likely to deliver more benefit than the control among a whole community. However, the same treatment may affect different individuals very differently. Therefore, a substantial amount of works focus on analyzing heterogeneity in treatment effects, of which the term refers to variation in the effects of treatment across individuals. This variation may provide theoretical insights, revealing how the effect of interventions depends on participants’ characteristics or how varying features of a treatment alters the effect of an intervention.

In this paper, we follow the potential outcome framework for causal inference \citep{Neyman23, Rubin74}, where each unit is assigned into either the treatment or the control group. Each unit has an observed outcome variable with a set of covariates. In randomized experiments and observational studies, it is desirable to replicate a sample as closely as possible by obtaining subjects from the treatment and control groups with similar covariate distributions when estimating causal effects. However, it is almost impossible to match observations exactly the same in both treatment and control groups in observational studies. To address this problem, it is usually preferred to define prespecified subgroups under certain conditions and estimate the treatment effects varying among subgroups. Accordingly, the conditional average treatment effect (CATE; \cite{Hahn98}) function is designed to capture heterogeneity of a treatment effect across subpopulations. The function is often conditioned on some component(s) of the covariates or a single statistic, like propensity score \citep{Rosenbaum83} and prognostic score \citep{Hansen08}. Propensity scores are the probabilities of receiving the treatment of interest; prognostic scores model the potential outcome under a control group assignment. To understand treatment effect heterogeneity in terms of propensity and prognostic scores, we assume that equal or similar treatment effects are observed along some intervals of the two scores. 

We target at constructing an estimator for treatment effects that conditions on only two quantities, propensity and prognostic scores, and we assume a piecewise constant structure in treatment effects. We take a step further from score-based matching algorithms and propose a data-driven approach that integrates the joint use of propensity and prognostic scores in a matching algorithm and a partition over the entire population via a non-parametric regression tree. In the first step, we estimate propensity scores and prognostic scores for each observed unit in the data. Secondly, we perform a $K$-nearest-neighbor matching of units of the treatment and control groups based on the two estimated scores and forth construct a proxy of individual treatment effects for all units. The last step involves growing a binary tree regressed on the two estimated scores. 

The complementary nature of propensity and prognostic score methods supports that conditioning on both the propensity and prognostic scores has the potential to reduce bias and improve the precision of treatment effect estimates, and it is affirmed in the simulation studies by \citet{Leacy14} and \citet{Antonelli18}. We also demonstrate such advantage for our proposed estimator across almost all scenarios examined in the simulation experiments.

Besides high precision in estimation, our proposed estimator demonstrates its superiority over state-of-arts methods with a few attractive properties as follows:
\begin{itemize}
    \item The estimator is computationally efficient. Propensity and prognostic scores can be easily estimated through simple regression techniques. Our matching algorithm based on the two scores largely reduces dimensionality compared to full matching on the complete covariates. Moreover, growing a single regression tree saves much time over other tree-based estimation methods, such as BART \citep{Hahn20} and random forests \citep{Wager18, Athey19a}.
    \item Many previous works in subgroup analysis, such as \citet{Assmann00} and \citet{Abadie18}, set stratification on the sample with a fixed number of subgroups before estimating treatment effects. These approaches require a pre-determination on the number of subgroups contained in the data, and they inevitably introduce arbitrariness into the causal inference. In comparison, our proposed method simultaneously identifies the underlying subgroups in observations through binary split according to propensity and prognostic scores and provides a consequential estimation of treatment effects on each subgroups.
    \item Although random forests based methods \citep{Wager18, Athey19a} achieve great performance in minimizing bias in estimating treatment effects, these ensemble methods are often referred to as ``black boxes''. It is hard to capture the underlying reason why the collective decision with the high number of operations involved is made in their estimation process, especially in the case when there are many features considered in the model. On the contrary, our proposed method provides a straightforward and low-dimensional summary of treatment effects simply based on two scores that are relatively easy to estimate. As a result, given the covariates of an observation, one can easily deduce the positiveness and magnitude of its treatment effect according to its probability of treatment receipt and potential outcome following the structure of the regression tree without carrying out feature selection process. 
    \item Instead of relying on a single feature or a subset of features, identifying subgroups with the joint distribution of all features will be of particular interest to policymakers seeking to target policies on those most likely to benefit. Therefore, many existing researches focus on subgroup stratification based on a single index that combines baseline characteristics. For instance, \citet{Abadie18} utilizes potential outcomes without treatment (prognostic scores) for endogenous stratification, and \citet{Kent07} stratifies experimental subjects based on their predicted probability of certain risks (propensity scores). Our method can be considered as a natural extension of the two previous works, which allows researchers and policy makers to merge a large number of baseline characteristics into simply two indices and identify the subgroup who are most in need of help in a population.\\
\end{itemize}

We review relevant literature on matching algorithms and estimation of heterogeneous treatment effects in Section \ref{sec2}. In Section \ref{sec3}, we provide the theoretical framework and preliminaries for the causal inference model. We propose our method for estimation and prediction in Section \ref{sec4}. Section \ref{sec5} lists the results of numerical experiments on multiple simulated data sets and two real-world data sets, following with the comparison with state-of-the-art methods in existing literature and the discussion on policy implications under different realistic scenarios.

\section{Relevant literature}
\label{sec2}
Statistical analysis of causality can be dated back to \citet{Neyman23}. Causal inference can be viewed as an identification problem \citep{Keele15}, for which statisticians are dedicated to learn the true causality behind the data. In reality, however, we do not have enough information to determine the true value due to a limited number of observations for analysis. This problem is also summarized as a ``missing data'' problem \citep{Ding18}, which stems from the \textit{fundamental problem of causal inference} \citep{Holland86}, that is, for each unit at most one of the potential outcomes is observed. Importantly, the causal effect identification problem, especially for estimating treatment effects, can only be resolved through assumptions. Several key theoretical frameworks have been proposed over the past decades. The potential outcome framework by \citet{Rubin74}, often referred to as the Rubin Causal Model (RCM; \cite{Holland86}), is a common model of causality in statistics at the moment. \citet{Dawid00} develops a decision theoretic approach to causality that rejects counterfactuals. \citet{Pearl95, Pearl09} advocate for a model of causality based on non-parametric structural equations and path diagrams.

\subsection*{Matching} 
To tackle the ``missing data'' problem when estimating treatment effects in randomized experiments in practice, matching serves as a very powerful tool. The main goal of matching is to find matched groups with similar or balanced observed covariate distributions \citep{Stuart10}. The exact $K$-nearest-neighbor matching \citep{Rubin74} is one of the most common, and easiest to implement and understand methods; and ratio matching \citep{Smith97, Rubin00, Ming01}, which finds multiple good matches for each treated individual, performs well when there is a large number of control individuals. \citet{Rosenbaum89, Gu93, Zubizarreta12, Zubizarreta17} developed various optimal matching algorithms to minimize the total sum of distances between treated units and matched controls in a global sense. \citet{Abadie06} studied the consistency of covariate matching estimators under large sample assumptions. Instead of greedy matching on the entire covariates, propensity score matching (PSM) by \citet{Rubin96} is an alternative algorithm that does not guarantee optimal balance among covariates and reduces dimension sufficiently. \citet{Imbens04} improved propensity score matching with regression adjustment. The additional matching on prognostic factors in propensity score matching was first considered by \citet{Rubin00}. Later, \citet{Leacy14} demonstrated the superiority of the joint use of propensity and prognostic scores in matching over single-score based matching in low-dimensional settings through extensive simulation studies. \citet{Antonelli18} extended the method to fit to high-dimensional settings and derived asymptotic results for the so-called doubly robust matching estimators. The sequential works by \citet{Aikens20} and \citet{Aikens22} pioneered in visualizing the relationship between propensity and prognostic scores as auxiliary in the estimation of treatment effects.

\subsection*{Subclassification} 
To understand heterogeneity of treatment effects in the data, subclassification, first used in \citet{Cochran68}, is another important research problem. The key idea is to form subgroups over the entire population based on characteristics that are either immutable or observed before randomization. \citet{Rosenbaum83, Rosenbaum85},  and \citet{Lunceford04} examined how creating a fixed number of subclasses according to propensity scores removes the bias in the estimated treatment effects, and \citet{Yang16} developed a similar methodology in settings with more than two treatment levels. Full matching \citep{Rosenbaum91, Hansen04, Stuart08} is a more sophisticated form of subclassification that selects the number of subclasses automatically by creating a series of matched sets. \citet{Schou15} presented three measures derived using the theory of order statistics to claim heterogeneity of treatment effect across subgroups. \citet{Su09} pioneered in exploiting standard regression tree methods \citep{Breiman84} in subgroup treatment effect analysis. Further, \citet{Athey16} derived a recursive partition of the population according to treatment effect heterogeneity. \citet{Hill11} was the first work to advocate for the use of Bayesian additive regression tree models (BART; \cite{Chipman10}) for estimating heterogeneous treatment effects, followed by a significant number of research papers focusing on the seminal methodology, including \citet{Green12}, \citet{Hill13} and \citet{Hahn20}. \citet{Abadie18} introduced endogenous stratification to estimate subgroup effects for a fixed number of subgroups based on certain quantiles of the prognostic score. More recently, \citet{Padilla21} combined the fused lasso estimator with score matching methods to lead to a data-adaptive subgroup effects estimator.

\subsection*{Machine Learning for Causal Inference}
For the goal of analyzing treatment effect heterogeneity, supervised machine learning methods play an important role. One of the more common ways for accurate estimation with experimental and observational data is to apply regression \citep{Imbens15} or tree-based methods \citep{Imai11}. From a Bayesian perspective, \citet{Heckman14} provided a principled way of adding priors to regression models, and \citet{Taddy16} developed Bayesian non-parametric approaches for both linear regression and tree models. The recent breakthrough work by \citet{Wager18} proposed the causal forest estimator arising from random forests from \citet{Breiman01}. More recently, \citet{Athey19a} took a step forward and enhanced the previous estimator based on generalized random forests. \citet{Imai13} adapted an estimator from the Support Vector Machine (SVM) classifier with hinge loss \citep{Wahba02}. \citet{Bloniarz16} studied treatment effect estimators with lasso regularization \citep{Tibshirani96} when the number of covariates is large, and \citet{Koch18} applied group lasso for simultaneous covariate selection and robust estimation of causal effects. In the meantime, a series of papers including \citet{Qian11}, \citet{Kunzel19} and \citet{Syrgkanis19}, focused on developing meta-learners for heterogeneous treatment effects that can take advantage of various machine learning algorithms and data structures.

\subsection*{Applied Work}
On the application side, the estimation of heterogeneous treatment effects is particularly an intriguing topic in causal inference with broad applications in scientific research. \citet{Gaines11} estimated heterogeneous treatment effects in randomized experiments in the context of political science. \citet{Dehejia02} explored the use of propensity score matching for nonexperimental causal studies with application in economics. \citet{Dahabreh16} investigated heterogeneous treatment effects to provide the evidence base for precision medicine and patient-centred care. \citet{Zhang17} proposed the Survival Causal Tree (SCT) method to discover patient subgroups with heterogeneous treatment effects from censored observational data. \citet{Rekkas20} examined three classes of approaches to identify heterogeneity of treatment effect within a randomized clinical trial, and \citet{Tanniou17} rendered a subgroup treatment estimate for drug trials.

\section{Preliminaries}
\label{sec3}
Before we introduce our method, we need to provide some mathematical background for treatment effect estimation. We follow Rubin's framework on causal inference \citep{Rubin74}, and assume a superpopulation or distribution $\mathcal{P}$ from which a realization of $n$ independent random variables is given as the training data. That is, we are given  $\{(Y_i(0), Y_i(1), X_i, Z_i)\}_{i=1}^n$ independent copies of $(Y(1),Y(0),X,Z)$, where $X_i \in \mathbb{R}^d$ is a $d$-dimensional covariate or feature vector, $Z_i \in \{0, 1\}$ is the treatment-assignment indicator, $Y_i(0) \in \mathbb{R}$ is the potential outcome of unit $i$ when $i$ is assigned to the control group, and $Y_i(1)$ is the potential outcome when $i$ is assigned to the treatment group.

One important and commonly used measure of causality in a binary treatment model is the average treatment effect (ATE; \cite{Imbens04}), that is, the mean outcome difference between the treatment and control groups. Formally, we write the ATE as
$$\text{ATE} := \mathbb{E}[Y(1)- Y(0)].$$
With the $n$ units in the study, we further define the individual treatment effect (ITE) of unit $i$ denoted by $D_i$ as
$$D_i:= Y_i(1)-Y_i(0).$$
Then, an unbiased estimate of the ATE is the sample average treatment effect 
$$\bar{Y}(1) - \bar{Y}(0) = \frac{1}{n}\sum_{i=1}^n D_i.$$
However, we cannot observe $D_i$ for any unit because a unit is either in the treatment group or in the control group, but not in both.

To analyze heterogeneous treatment effects, it is natural to divide the data into subgroups (e.g., by gender, or by race), and investigate if the average treatment effects are different across subgroups. Therefore, instead of estimating the ATE or the ITE directly, statisticians seek to estimate the conditional average treatment effect (CATE), defined by
\begin{equation}
    \tau(x) := \mathbb{E}[Y(1) - Y(0) \ | \ X = x].
\end{equation}
The CATE can be viewed as an ATE in a subpopulation defined by $\{X=x\}$, i.e. the ATE conditioned on membership in the subgroup.

We also recall the propensity score \citep{Rosenbaum83}, denoted by $e(X)$, and defined as
$$e(X) = \mathbb{P}(Z = 1 \ | \ X).$$
Thus, $e(X)$ is the the probability of receiving treatment for  
a unit with covariate $X$. In addition, we consider prognostic scores, denoted by $p(X)$, for potential outcomes. The prognostic score is first defined by \citet{Hansen08} as any quantity $p(X)$ that satisfies
$$Y(0) \indep X \ | \ p(X),$$
and in this manuscript, we use the conventional definition as the predicted outcome under the control condition \citep{Aikens20}: 
$$p(X) = E[Y(0) \ | \ X].$$
We are interested in constructing a 2d summary of treatment effects based on propensity and prognostic scores. Instead of conditioning on the entire covariates or a subset of it in the CATE function, we express our estimand, named as scored-based subgroup CATE, by conditioning on the two scores:
\begin{equation}
\label{estimand}
\tau(x):=\mathbb{E}[Y(1)-Y(0) \ | \ e = e(x), p = p(x)].
\end{equation}
For interpretability, we assume that treatment effects are piecewise constant over a 2d grid of propensity and prognostic scores. Specifically, there exists a partition of intervals $\{I_1^e, ..., I_s^e\}$ of $[0,1]$ and another partition of intervals  $\{I_1^p, ..., I_t^p\}$ of $\mathbb{R}$ such that for any $i \in \{1,...,s\}$ and $j \in \{1,...,t\}$, we have  
$$\tau(x) \equiv C_{i,j} \ \ \text{ for } x  \text{ s.t. } e(x) \in I^e_i, p(x) \in I^p_j,$$
where $C_{i,j} \in \mathbb{R}$ is a constant.

Moreover, our estimation of treatment effects relies on the following assumptions:
\begin{assumption}
\label{as1}
Throughout the paper, we maintain the Stable Unit Treatment Value Assumption (SUTVA; \cite{Imbens15}), which consists of two components: no interference and no hidden variations of treatment. Mathematically, for unit $i =1,..., n$ with outcome $Y_i$ and treatment indicator $Z_i$, it holds that
$$Y_i(Z_1, Z_2, ..., Z_n) = Y_i(Z_i).$$
\end{assumption}
Thus, the SUTVA requires that the potential outcomes of one unit should be unaffected by the particular assignment of treatments to the other units. Furthermore, for each unit, there are no different forms or versions of each treatment level, which lead to different potential outcomes.

\begin{assumption}
\label{as2}
The assumption of probabilistic assignment holds. This  requires the assignment mechanism to imply a non-zero probability for each treatment value, for every unit. For the given covariates $X$ and treatment-assignment indicator $Z$, we must have
$$0 < \mathbb{P}(Z = 1 \ | \ X) < 1,$$
almost surely.
\end{assumption}
This condition, regarding the joint distribution of treatments and covariates, is also known as overlap in some literature (see Assumption 2.2 in \cite{Imbens04} and \cite{DAmour21}), and  it is necessary for estimating treatment effects everywhere in the defined covariate space. Note that $\mathbb{P}(Z_i = 1 \ | \ X_i)$ is  the propensity score. In other words, Assumption \ref{as2} requires that the propensity score, for all values of the treatment and all combinations of values of the confounders, be strictly between 0 and 1.

\begin{assumption}
\label{as3}
We make the assumption that
$$(Y(0), Y(1)) \indep Z \ | \ e(X), p(X)$$
holds.
\end{assumption}
This assumption is an implication of the usual unconfoundedness assumption:
\begin{equation}
\label{confound}
    (Y(0), Y(1)) \indep Z \ | \ X
\end{equation}
Combining Assumption \ref{as2} and that in Equation (\ref{confound}), the conditions are typically referred as \textit{strong ignorability} defined in \citet{Rosenbaum83}. Strong ignorability states which outcomes are observed or missing is independent of the missing data conditional on the observed data. It allows statisticians to address the challenge that the ``ground truth'' for the causal effect is not observed for any individual unit. We rewrite the conventional assumption by replacing the vector of covariates  $x$ with the joint of propensity score $e(x)$ and $p(x)$ to accord with our estimation target.

Provided that Assumptions \ref{as1}-\ref{as3} hold, it follows that $$\mathbb{E}[Y(z) \ | \ e = e(x), p = p(x)] = \mathbb{E}[Y \ | \ e = e(x), p = p(x), Z = z],$$ and thus our estimand (\ref{estimand}) is equivalent to 
\begin{equation}
    \label{eqn:final-estimand}
    \tau(x)=\mathbb{E}[Y \ | \ e = e(x), p = p(x), Z = 1] - \mathbb{E}[Y \ | \ e = e(x), p = p(x), Z = 0].
\end{equation}

Thus, in this paper we focus on estimating (\ref{eqn:final-estimand}), which is equivalent to 
(\ref{estimand}) if the assumptions above hold, but might be different if Assumption \ref{as3} is violated.

\section{Methodology}
\label{sec4}
We now formally introduce our proposal of a three-step method for estimating heterogeneous treatment effects and the estimation rule for a given new observation. We assume  a sample of size $n$ with covariate $X$, treatment indicator $Z$, and outcome variable $Y$, where the notations inherit from the previous section. Generally, we consider a low-dimensional set-up, where the sample size $n$ is larger than the covariate dimension $d$. An extension of our proposed method to the high-dimensional case is discussed in this section as well.

\subsection*{Step 1}
We first estimate propensity and prognostic scores for all observations in the sample. For propensity score, we apply a logistic regression on the entire covariate $X$ and the treatment indicator $Z$ by solving the optimization problem
\begin{equation}
\label{eq:alpha}
    \hat\alpha = \underset{\alpha \in \mathbb{R}^d}{\arg\min} \ -\sum_{i=1}^n \Big[Z_i\log\sigma(X_i^\top\alpha)+(1-Z_i)\log(1-\sigma(X_i^\top\alpha))\Big],
\end{equation}
where $\sigma(x)= \frac{1}{1+\exp(-x)}$ is the logistic function. With the coefficient vector $\hat\alpha$, we compute the estimated propensity scores $\hat e$ by 
$$\hat e_i = \sigma\left(X_i^\top \hat\alpha\right).$$

For prognostic score, we restrict to the controlled group, and regress the outcome variable $Y$ on the covariate $X$ through ordinary least squares: we solve
\begin{equation}
\label{eq:theta}
\hat\theta = \underset{\theta \in \mathbb{R}^d}{\arg\min} \ \sum_{i: Z_i = 0}(Y_i-X_i^\top\theta)^2,
\end{equation}
and we estimate prognostic scores as
$$\hat p_i = X_i^\top \hat\theta.$$

\subsection*{Step 2}
Next, we perform a nearest-neighbor matching based on the two estimated scores from the previous step. We adapt the notation from \citet{Abadie06}, and use the Mahalanobis distance norm as the distance metric in the matching algorithm. Mahalanobis distance matching, first proposed by \citet{Hansen06}, has been widely used in matching algorithms for causal inference \citep{Leacy14, Aikens20}. It has been found to perform well when the dimension of covariates are relatively low \citep{Stuart10}, and thus it is more suitable than using standard Euclidean distance in our case when we have only propensity scores and prognostic scores as covariates.

Formally, for the units $i$ and $j$ with estimated propensity scores $\hat e_i, \hat e_j$ and propensity scores $\hat p_i, \hat p_j$, we define the score-based Mahalanobis distance between $i$ and $j$ by
$$d(i,j) = \left[\begin{pmatrix}
\hat e_i\\
\hat p_i
\end{pmatrix} - \begin{pmatrix}
\hat e_j\\
\hat p_j
\end{pmatrix}\right]^\top \Sigma^{-1}\left[\begin{pmatrix}
\hat e_i\\
\hat p_i
\end{pmatrix} - \begin{pmatrix}
\hat e_j\\
\hat p_j
\end{pmatrix}\right],$$
where $\Sigma$ denotes the variance-covariance matrix of $(\hat e, \hat p)^\top$.

Let $j_k(i)$ be the index $j \in \{1,2,...,n\}$ that solves $Z_j = 1 - Z_i$ and 
$$\sum_{l: Z_l = 1 - Z_i} \textbf{1}\left\{d(l,i) \leq d(j,i)\right\} = k,$$
where $\textbf{1}\{\cdot\}$ is the indicator function. This is the index of the unit that is the $k$th closest to unit $i$ in terms of the distance between two scores, among the units with the treatment opposite to that of unit $i$. We can now construct the $K$-nearest-neighbor set for unit $i$ by the set of indices for the first $K$ matches for unit $i$, $$\mathcal{J}_K(i) = \{j_1(i),...,j_K(i)\}.$$
We then compute
\begin{equation}
\label{tilde}
    \tilde Y_i = (2Z_i-1)\left(Y_i - \frac{1}{K}\sum_{j \in \mathcal{J}_K(i)} Y_{j}\right).
\end{equation}
Intuitively, the construction of $\tilde Y$ gives a proxy of the individual treatment effect (ITE) on each unit. We find $K$ matches for each unit in the opposite treatment group based on the similarity of their propensity and prognostics scores, and the mean of the $K$ matches is used to estimate the unobserved potential outcome for each unit.

\subsection*{Step 3}
\begin{figure}[ht!]
\centering
\begin{subfigure}{.45\textwidth}
  \centering
  \includegraphics[width=.9\linewidth]{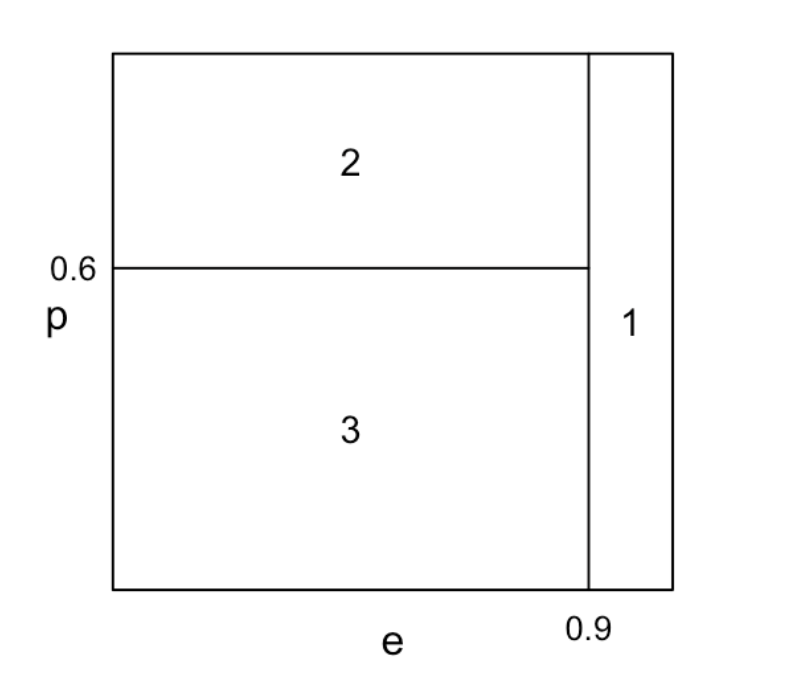}
\end{subfigure}%
\begin{subfigure}{.45\textwidth}
  \centering
  \includegraphics[width=.9\linewidth]{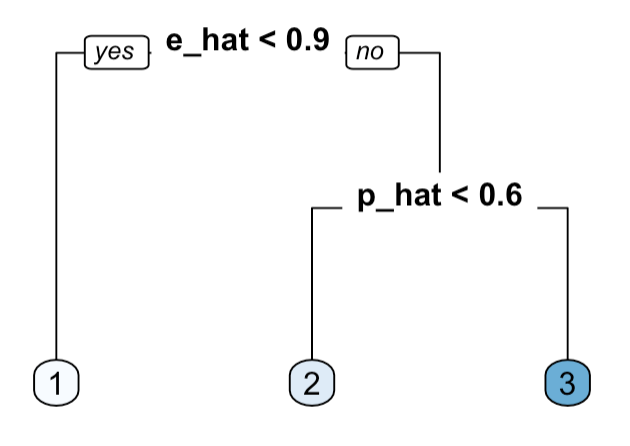}
\end{subfigure}
\captionsetup{justification=centering}
\caption{\textit{Left: } a hypothetical partition over the 2d space of propensity and prognostic scores with the true values of piecewise constant treatment effects; \textit{Right:} a sample regression tree $T$ constructed in Step 3.}
\label{fig1}
\end{figure}

The last step involves denoising of the point estimates of the individual treatment effects $\tilde Y$ obtained from Step 2. The goal is to partition all units into subgroups such that the estimated treatment effects would be constant over some 2d intervals of propensity and prognostic scores (see the left of Figure \ref{fig1}).

To perform such stratification, we grow a regression tree on $\tilde Y$, denoted by $T$, and the regressors are the estimated propensity scores $\hat e$ and the estimated prognostic scores $\hat p$ from Step 1. We follow the very general rule of binary recursive partitioning to build the tree $T$: allocate the data into the first two branches, using every possible binary split on every covariate; select the split that minimizes Gini impurity, continue the optimal splits over each branch along the covariate’s values until the minimum node size is reached. To avoid overfitting, we set the minimum node size as 20 in our model. Choosing other criteria such as information gain instead of Gini impurity is another option for splitting criteria. A 10-fold cross validation is also performed at meantime to prune the large tree $T$ for deciding the value of cost complexity. Cost complexity is the minimum improvement in the model needed at each node. The pruning rule follows that if one split does not improve the overall error of the model by the chosen cost complexity, then that split is decreed to be not worth pursuing (see more details in Section 9.2 of \cite{Hastie01}).

The final tree $T$ (see the right plot of Figure \ref{fig1}) contains a few terminal nodes, and these are the predicted treatment effects for all units in the data. The values exactly represent a piecewise constant stratification over the 2d space of propensity and prognostic scores.

\subsection*{Estimation on a new unit}
After we obtain the regression tree model $T$ in Step 3, we can now estimate the value of the individual treatment effect corresponding to a new unit with covariate $x_{\text{new}}$. 

We first compute the estimated propensity and prognostic scores for the new observation by
$$\hat e_{\text{new}} = \sigma(x_{\text{new}}^\top \hat \alpha), \ \hat p_{\text{new}} = x_{\text{new}}^\top \hat \theta,$$
where $\hat\alpha$ and $\hat\theta$ are the solutions to Equations (\ref{eq:alpha}) and (\ref{eq:theta}) respectively. Then with the estimated propensity score $\hat e_{\text{new}}$ and prognostic score $\hat p_{\text{new}}$, we can get an estimate of the treatment effect for this unit following the binary predictive rules in the tree $T$.

\subsection*{High-Dimensional Estimator}
In a high-dimensional setting where the covariate dimension $d$ is much larger than the sample size $n$, we can estimate the propensity and prognostic scores by adding a lasso ($l1$-based) penalty \citep{Tibshirani96} instead. This strategy was first proposed and named as ``doubly robust matching estimators'' (DRME) by \citet{Antonelli18}. The corresponding optimization problems for the two scores can be written as
\begin{align*}
    \hat\alpha &= \underset{\alpha \in \mathbb{R}^d}{\arg\min} \ -\sum_{i=1}^n \Big[Y_i\log\sigma(X_i^\top\alpha)+(1-Y_i)\log(1-\sigma(X_i^\top\alpha))\Big]+ \lambda_1\sum_{j=1}^d|\alpha_j|,\\
\hat\theta &= \underset{\theta \in \mathbb{R}^d}{\arg\min} \ \sum_{i: Z_i = 0}(Y_i-X_i^\top\theta)^2 + \lambda_2\sum_{j=1}^d|\theta_j|.
\end{align*}
The selection of the tuning parameter $\lambda_1, \lambda_2$ can be determined by any information criteria (AIC, BIC, and etc.). In practice, we use $10$-fold cross-validation (CV) to select the value of $\lambda$. Then we perform the $K$-nearest-neighbor matching based on propensity and prognostic scores to get the estimates of individual treatment effects using Equation (\ref{tilde}).

We extend the above estimator with our proposal of applying a regression tree on the estimated propensity and prognostic scores. The procedure for estimation of subgroup heterogeneous treatment effects and estimation on a new unit remains the same as Step 3 for low-dimensional set-ups.\\

\begin{remark}
The choice of the number of nearest neighbors is a challenging problem. Updating distance metrics for every observation is computationally expensive, and choosing a value that is too small leads to a higher influence of noise on estimation. With regards to the application of nearest neighbor matching in causal inference, \citet{Abadie06} derived large sample properties of matching estimators of average treatment effects with a fixed number of nearest neighbors, but the authors did not provide any details on how to select the exact number of neighbors. Conventional settings on the number of nearest neighbors in current literature is to set $K = 1$ (one-to-one matching; \citep{Stuart10, Austin16}). However, \citet{Ming00} suggested that in observational studies, substantially greater bias reduction is possible through matching with a variable number of controls rather than exact pair matching. 

In Appendix \ref{appA}, we conduct a simulation study following one of the generative models from Section \ref{sec5} to show how sensitive estimation accuracy is to the number of nearest neighbors selected and setting $K$ to a large number other than 1 is more `sensible' to reduce estimation bias. Although it is usually difficult to select a perfect value of $K$ in practice, simply setting $K \approx \log(n)$ as suggested by \citet{Brito97} leads to reasonable results for a data sample of size $n$. Throughout all our experiments in the next section, setting $K$ to the integer closest to the value $\log(n)$ provides estimates with high accuracy and does not require too much computational cost.
\end{remark}

\subsection*{Computational Complexity}
Our method is composed of three steps as introduced above. We first need to implement a logistic regression for estimating propensity score for a sample with size $n$ and ambient dimension $d$. To solve a logistic regression optimization problem involves using the proximal Newton method, and its computational complexity is of $\#\text{ of iterations} \cdot O(nd)$ \citep{Yuan12}. In general, the algorithm converges after a small number of iterations, and we can safely assume it is of a constant order. As a result, the overall time complexity of solving a logistic regression is $O(nd)$. The estimation of prognostic scores also requires a computational cost of $O(nd)$, and for high-dimensional settings, the cost of solving the solution via coordinate descent remains at $O(nd)$ \citep{Friedman10} . The complexity of a $K$-nearest-neighbor matching based on the two estimated scores in the second step is of $O(Kn)$ \citep{von07}, and the selection of $K \approx \log(n)$ leads to a complexity of $O(n\log n)$. In the third step, we grow a regression tree based on two estimated scores, and it requires a computational complexity of $O(n\log n)$. 

The overall computational complexity of our method depends on the comparison between the order of $d$ and $\log(n)$. Our method attains a computational complexity of $O(nd)$ if the order of $d$ is greater than that of $\log(n)$. Otherwise, the complexity becomes $O(n\log n)$. 

\section{Experiments}
\label{sec5}
In this section, we will examine the performance of our proposed estimator (PP) in a variety of simulated and real data sets. The baseline estimators we compete against are leave-one-out endogenous stratification (ST; \cite{Abadie18}), causal forest (CF; \cite{Wager18}), single-score matching including propensity-score matching (PSM) and prognostic-score matching. Note that in the original research by \citet{Abadie18}, the authors restricted their attention to randomized experiments, because this is the setting where endogenous stratification is typically used. However, they mentioned the possibility of applying the method on observational studies. We take this into consideration, and make their method as one of our competitors.

We implement our methods in R, using the packages ``MatchIt'' for $K$-nearest-neighbor Mahalanobis distance matching and ``caret'' for growing a non-parametric regression tree with cross validation over complexity parameter. Throughout, we set the number of nearest neighbors, $K$, to be the closest integer to $\log(n)$, where $n$ is the sample size. For causal forest, we directly use the R package ``grf'' developed by \citet{Athey19a}, following with an automatic cross-validation to select values of hyper-parameters \citep{Athey19b} and a $K$-fold cross-fitting under a conventional setting of $K=10$ recommended by \citet{Nie21}. Software that replicate all the simulations is available on the authors' Github page.

We evaluate the performance of each method according to two aspects, accuracy and uncertainty quantification. The results for single-score matching algorithms are not reported in this paper because of very poor performance throughout all scenarios.

\subsection{Simulated Data} 
We first examine on the following simulated data sets under six different data generation mechanisms. We get insights from the simulation study in \citet{Leacy14} for the models considered in Scenarios 1-3. The propensity score and outcome (prognosis) models in Scenarios 1 and 3 are characterized by additivity and linearity (main effects only), but with different piecewise constant structures in the true treatment effects over a 2d grid of the two scores. We add non-additivity and non-linear terms to both propensity and prognosis models in Scenarios 2. In other words, both propensity and prognostic scores are expected to be misspecified in these two models if we apply generalized linear models directly in estimation. Scenario 4 comes from \citet{Abadie18}, with a constant treatment effect over all observations. Scenario 5 is modified from \citet{Wager18} (see Equation 27 there), in which the propensity model follows a continuous distribution instead of a linear structure. In addition, the true treatment effect is not a function of propensity and prognostic scores. In Scenario 6, we break the mold assumption of nicely delineated subgroups in treatment effects over 2d space of the two scores, and use a smooth function instead. A high-dimensional setting ($d >> n$) is examined in Scenario 7, where the generative model inherits from Scenario 1. We also include a table summarizing all simulation scenarios in Appendix \ref{appB} to help the audience interpret the result of the simulation study more readily.

We first introduce some notations used in the experiments: the sample size $n$, the ambient dimension $d$, as well as the following functions:
\begin{align*}
   &\text{true treatment effect: } \tau^*(X) = \mathbb{E}\left[Y(1) - Y(0)\big|X\right],\\
   &\text{treatment propensity: } e(x) = \mathbb{P}(Z=1|X=x),\\
   &\text{treatment logit: } \text{logit}(e(x)) = \log\left(\frac{e(x)}{1-e(x)}\right).
\end{align*}
Throughout all the models we consider, we maintain the unconfoundedness assumption discussed in Section \ref{sec3}, generate the covariate $X$ following a certain distribution, and entail homoscedastic Gaussian noise $\epsilon$. 

We evaluate the accuracy of an estimator $\tau(X)$ by the mean-squared error for estimating $\tau^*(X)$ at a random example $X$, with $m=n/10$, defined by
$$\text{MSE}(\hat\tau(X)) := \frac{1}{m}\sum_{i=1}^m \left[\hat\tau_i(X)-\tau_i^*(X)\right]^2.$$ 
We record the averaged MSE over 1000 Monte Carlo trials for each scenario. For each Monte Carlo trial, we use train-test split to evaluate model performance. In detail, we first obtain $n$ training sample and train models on this set. We then generate a separate testing sample under the same data generation mechanism with sample size $n/10$. We make predictions on the test data using the trained models and compute the MSEs from different models accordingly. In terms of uncertainty quantification, we measure the coverage probability of $\tau(X)$ with a target coverage rate of 0.95. For endogenous stratification and our proposed method, we use non-parametric bootstrap with replacement to construct the empirical quantiles for each unit. The details on the implementation of non-parametric bootstrap methods are presented in Appendix \ref{appC}. For causal forest, we construct 95$\%$ confidence intervals by estimating the standard errors of estimation using the ``grf'' package. We do not include the results for Scenario 7, a high-dimensional setting, because the asymptotic normality guarantees of non-parametric bootstrap and random forests are not valid in high dimensions \citep{Wager18, Karoui18}. \\

\noindent \textbf{Scenario 1.}
With $d \in \{2,10,50\}$, $n \in \{1000,5000\}$, for $i = 1,...,n$, we generate the data as follows:
\begin{align*} 
&Y_i = p(X_i) + Z_i\cdot\tau_i^* + \epsilon_i,\\
&\tau_i^* = \textbf{1}_{\{e(X_i) < 0.6, p(X_i) < 0\}},\\
&\text{logit}(e(X_i)) = X_i^\top\beta^e,\\ 
&p(X_i) = X_i^\top\beta^p,\\
&X_i \overset{i.i.d.}{\sim} \mathcal{U}[0,1]^d,\\
&\epsilon_i \overset{i.i.d.}{\sim} \mathcal{N}(0,1),
\end{align*}
where $\beta^e$ and $\beta^p$ are randomly generated from $\mathcal{U}[-1,1]^d$ in each iteration.
\\
\\
\noindent \textbf{Scenario 2.} We now add some interaction and non-linear terms to the propensity and prognostic models in Scenario 1, while keeping the set-ups of the covariate $X$, the response $Y$, the true treatment effect $\tau^*$ and the error term $\epsilon$ unchanged. We set $d = 10$ and $n = 5000$ in this case.
\begin{align*}
    \text{logit}(e(X_i)) &= X_i^\top \beta^e + 0.5 X_{i1}X_{i3} + 0.7X_{i2}X_{i4} + 0.5 X_{i3}X_{i5} \\
    & \,\,\,\ + 0.7 X_{i4}X_{i6} + 0.5 X_{i5}X_{i7} + 0.5X_{i1}X_{i6}\\
     & \,\,\,\ + 0.7 X_{i2}X_{i3} + 0.5 X_{i3}X_{i4} + 0.5 X_{i4}X_{i5}\\
     & \,\,\,\ + 0.5 X_{i5}X_{i6} + X_{i2}^2 + X_{i4}^2 - X_{i7}^2 \\
     p(X_i) &= X_i^\top \beta^p + 0.5 X_{i1}X_{i3} + 0.7X_{i2}X_{i4} + 0.5 X_{i3}X_{i8}, \\
    & \,\,\,\ + 0.7 X_{i4}X_{i9} + 0.5 X_{i8}X_{i10} + 0.5X_{i1}X_{i9}\\
     & \,\,\,\ + 0.7 X_{i2}X_{i3} + 0.5 X_{i3}X_{i4} + 0.5 X_{i4}X_{i8}\\
     & \,\,\,\ + 0.5 X_{i8}X_{i9} + X_{i3}^2 + X_{i4}^2 - X_{i10}^2.
\end{align*}
Again for each simulation, $\beta^e$ and $\beta^p$ are randomly generated from $\mathcal{U}[-1,1]^d$.
\\
\\
\textbf{Scenario 3.} In this case, we define the true treatment effect with a more complicated piecewise constant structure over the 2d grid, under the same model used in Scenario 1, with $d = 10$ and $n =5000$:
\begin{align*}
    \tau_i^* &= \begin{cases}
0 & \text{if } e(X_i) \leq 0.6, p(X_i)  \leq 0,\\
1 & \text{if } e(X_i) \leq 0.6, p(X_i) > 0\text{ or }  e(X_i) > 0.6, p(X_i) \leq 0,\\
2 & \text{if } e(X_i) > 0.6, p(X_i) > 0.
\end{cases}
\end{align*}
\\
\textbf{Scenario 4.} Setting $d = 10$ and $n = 5000$, the data is generated as:
\begin{align*}
    Y_i &= 1 + \beta^\top X_i +\epsilon_i,\\
    X_i &\overset{i.i.d.}{\sim} \mathcal{N}(0, \textbf{I}_{d\times d}),\\
    \epsilon_i &\overset{i.i.d.}{\sim}  \mathcal{N}(0, 100-d),
\end{align*}
where $\beta = (1,...,1)^\top \in \mathbb{R}^d$. Moreover, the treatment indicators for the simulations are such that $\sum_i Z_i = \lceil n/2\rceil$. By construction, the vector of treatment effects satisfies $\tau^* = 0$.
\\ \\
\noindent \textbf{Scenario 5.} The data satisfies
\begin{align*}
    \tau^*_i &= (X_{i1} + X_{i2}  + ... + X_{i10}) / 10,\\
    Y_i &= 2X_i^\top \textbf{e}_1 - 1 + Z_i\cdot\tau^*_i + \epsilon_i,\\
    Z_i &\sim \text{Binom}(1, e(X_i)),\\
    X_i &\overset{i.i.d.}{\sim} \mathcal{U}[0,1]^d ,\\
    e(X_i) &= \frac{1}{4}[1+\beta_{2,4}(X_i^\top \textbf{e}_1)],\\
    \epsilon_i &\overset{i.i.d.}{\sim}  \mathcal{N}(0, 1),
\end{align*}

\noindent where $\textbf{e}_1 = (1,0,...,0)$. We compare the results of different methods under the setting: $d = 10, n= 5000$. \\

\noindent \textbf{Scenario 6.}
For this scenario, we examine the performance when the treatment effect takes on a smooth interval of propensity and prognostic scores. We use the same model in Scenario 1 for generating the two scores with $d= 10$ and $n = 5000$, and we inherit and modify the generative function (28) from \citet{Wager18} for treatment effects as follows:
$$\tau^*_i = \zeta(e(X_i))\cdot\zeta(p(X_i)), \ \zeta(x) = 1 + \frac{1}{1 + e^{-20(x-1/3)}}.$$ 

\noindent \textbf{Scenario 7.} In the last case, we study the performance of different estimators on a high-dimensional data. The data model follows
\begin{align*}
&X_i \overset{i.i.d.}{\sim} \mathcal{U}[0,1]^d,\\
&Y_i = p(X_i) + Z_i\cdot\tau_i^* + \epsilon_i,\\
&\tau_i^* = \textbf{1}_{\{e(X_i) < 0.6, p(X_i) < 0\}},\\
&\text{logit}(e(X_i)) = 0.4X_{i1} +  0.9X_{i2} -0.4X_{i3} -0.7X_{i4}-0.3X_{i5} + 0.6X_{i6},\\
&p(X_i) = 0.9X_{i1} -0.9X_{i2} +  0.2 X_{i3} -0.2X_{i4} +0.9X_{i5} -0.9X_{i6},\\
&\epsilon_i \overset{i.i.d.}{\sim} \mathcal{N}(0,1).
\end{align*}
We select $n = 3000$ and $d = 5000$ for analysis.\\

\begin{figure}[H]
\centering
\begin{subfigure}{.45\textwidth}
  \centering
  \includegraphics[height = 36mm,width=.9\linewidth]{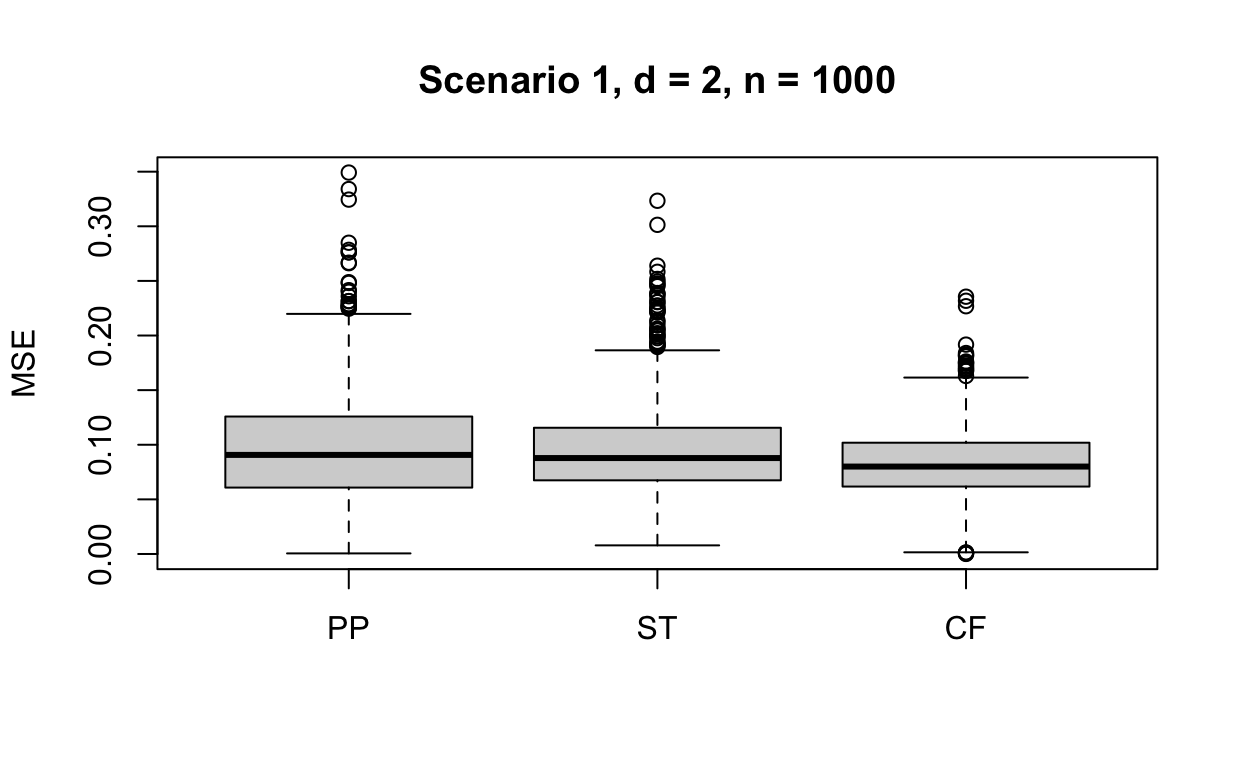}
\end{subfigure}%
\begin{subfigure}{.45\textwidth}
  \centering
  \includegraphics[height = 36mm,width=.9\linewidth]{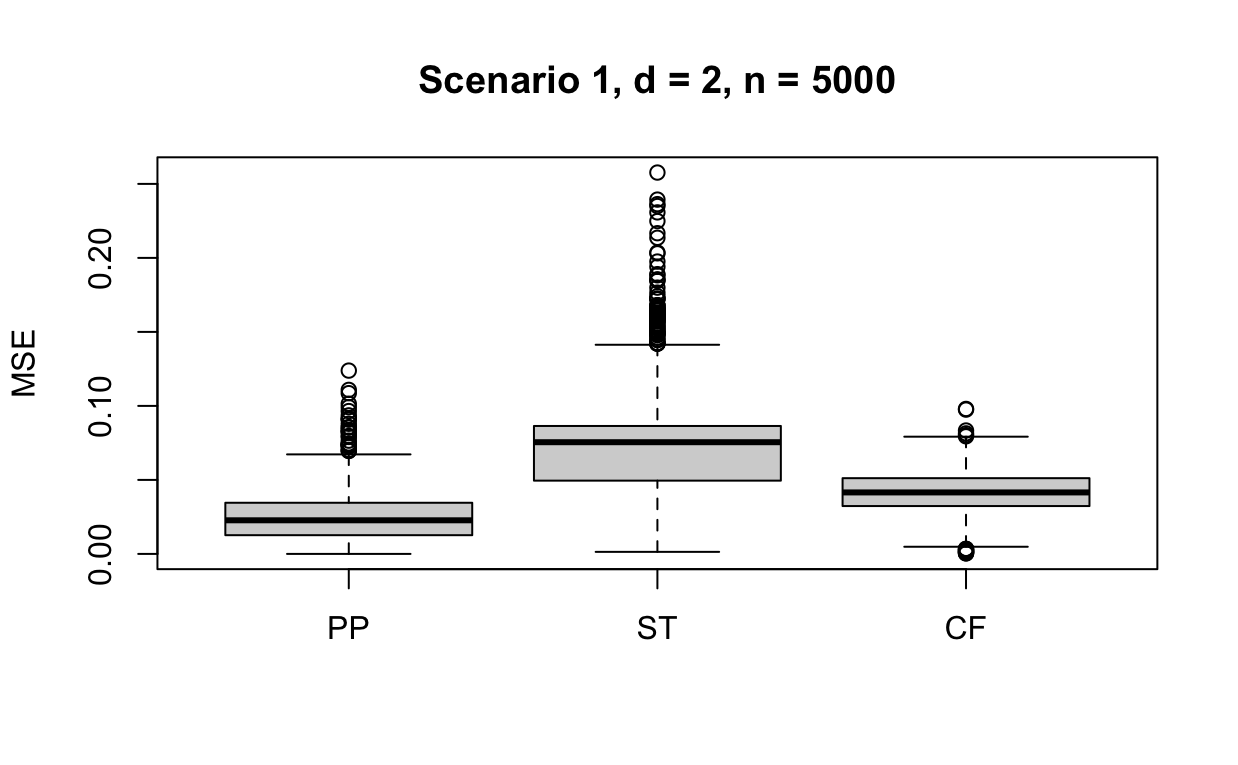}
\end{subfigure}\\
\begin{subfigure}{.45\textwidth}
  \centering
  \includegraphics[height = 36mm,width=.9\linewidth]{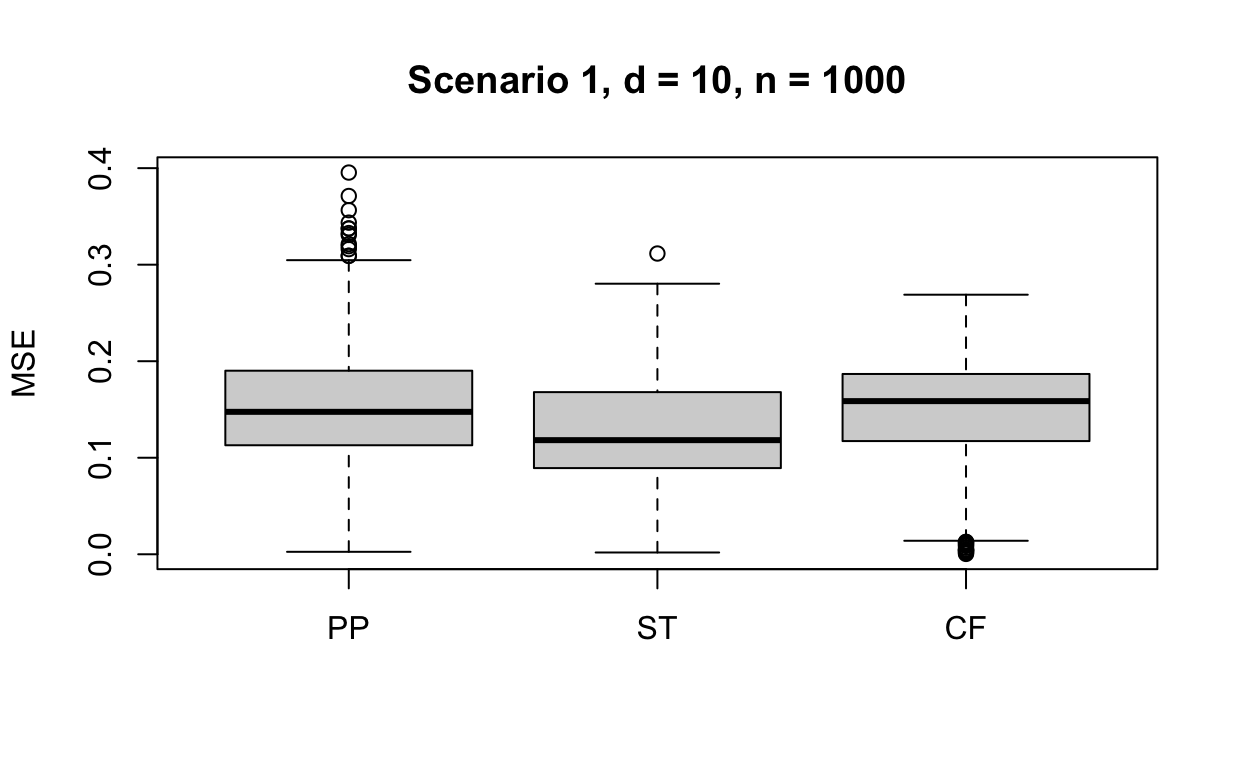}
\end{subfigure}%
\begin{subfigure}{.45\textwidth}
  \centering
  \includegraphics[height = 36mm,width=.9\linewidth]{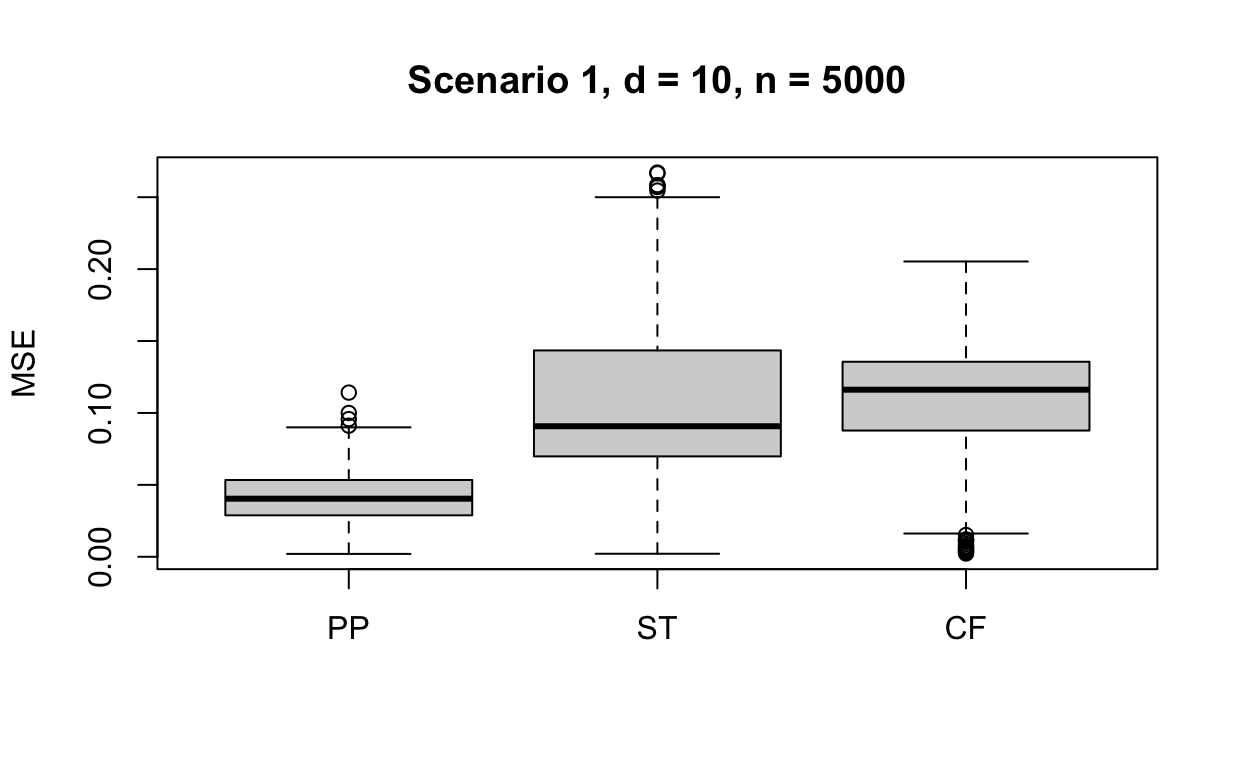}
\end{subfigure} \\
\begin{subfigure}{.45\textwidth}
  \centering
  \includegraphics[height = 36mm,width=.9\linewidth]{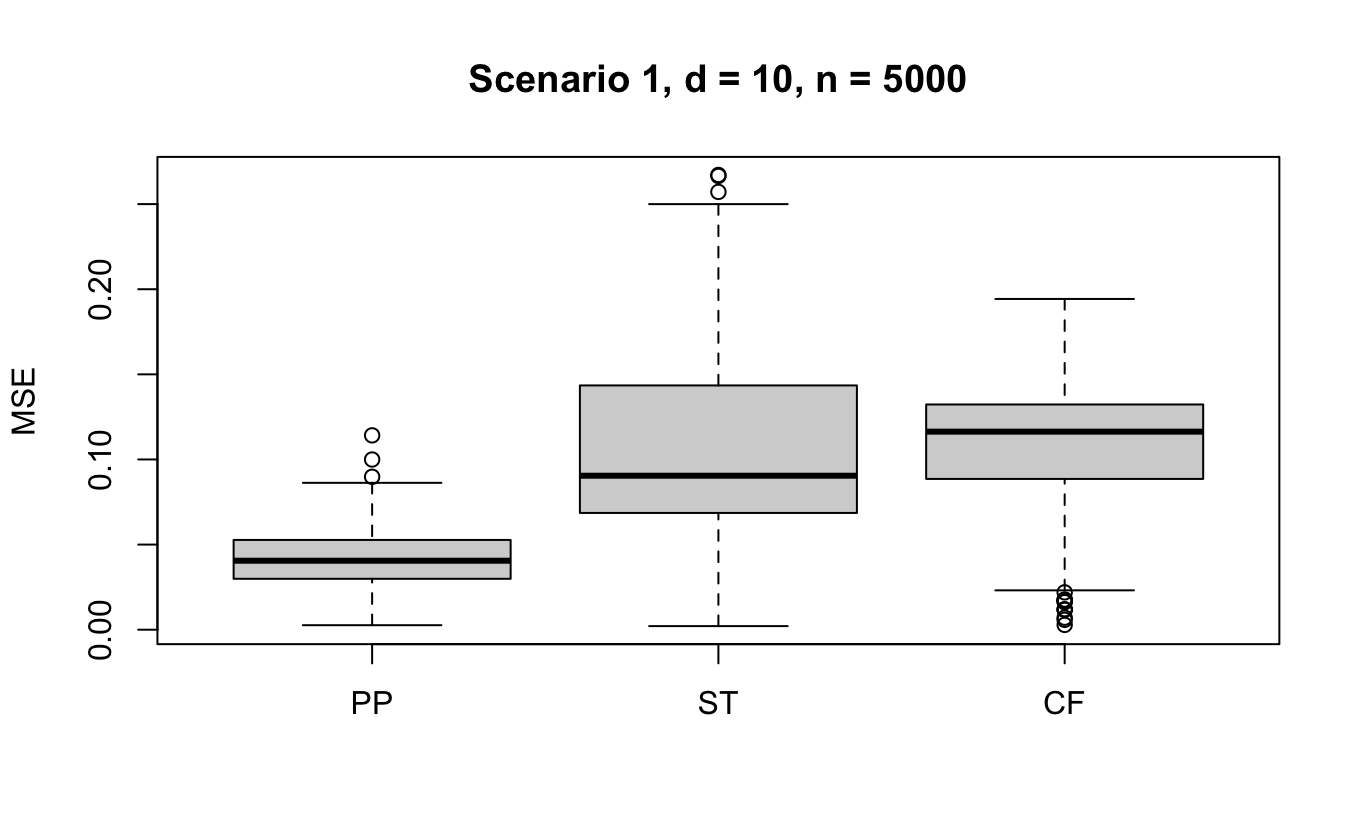}
\end{subfigure}%
\begin{subfigure}{.45\textwidth}
  \centering
  \includegraphics[height = 36mm,width=.9\linewidth]{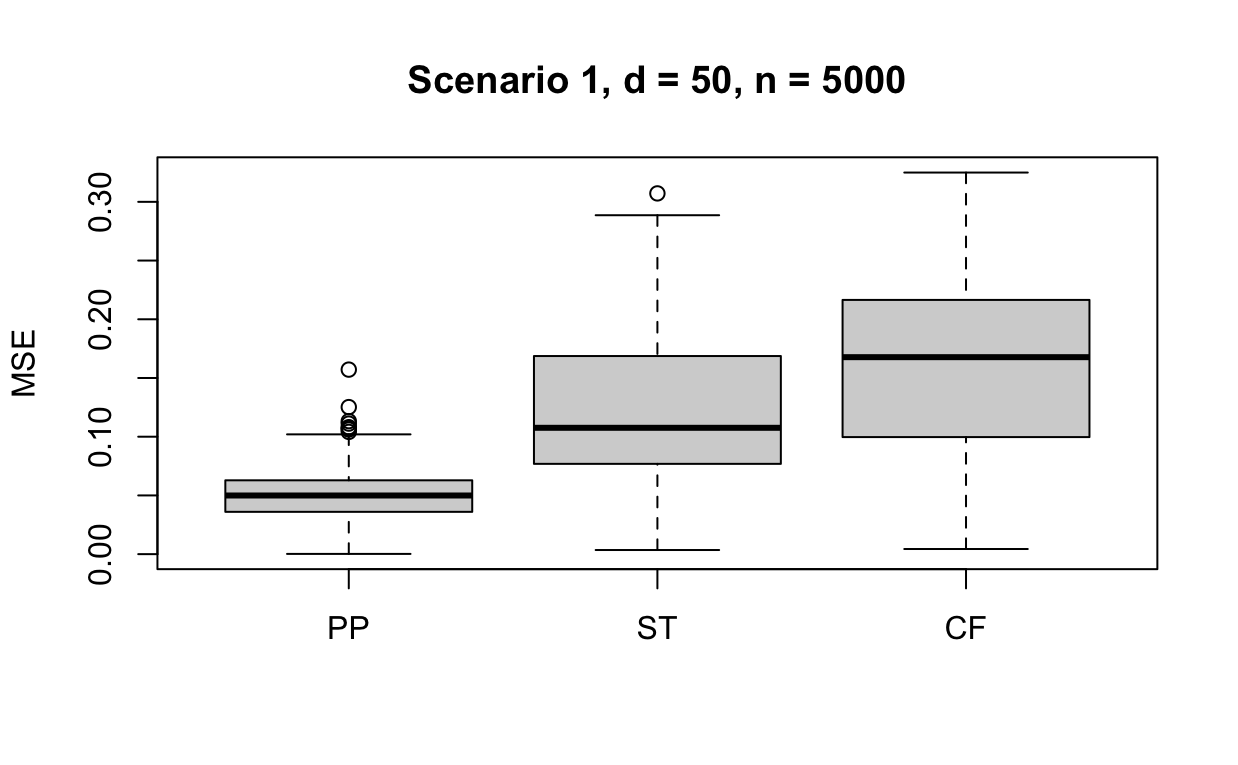}
\end{subfigure}\\
\begin{subfigure}{.45\textwidth}
  \centering
  \includegraphics[height = 36mm,width=.9\linewidth]{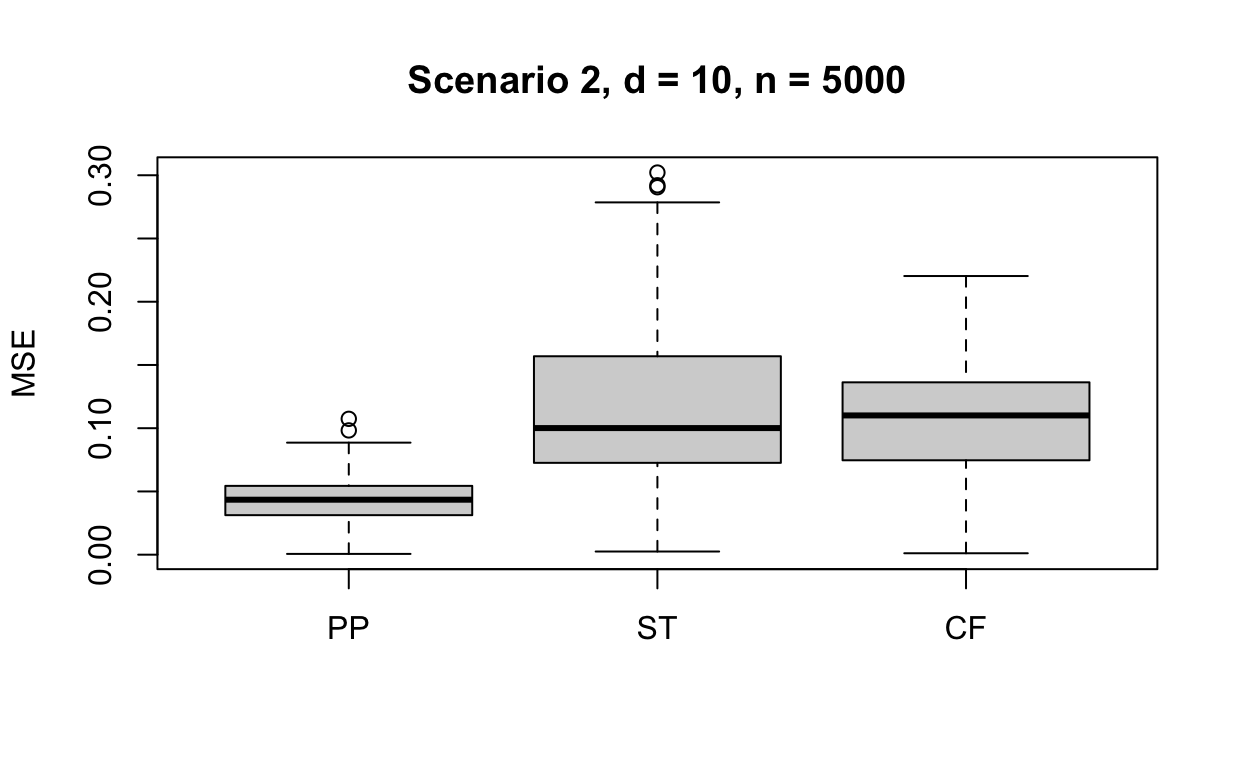}
\end{subfigure}%
\begin{subfigure}{.45\textwidth}
  \centering
  \includegraphics[height = 36mm,width=.9\linewidth]{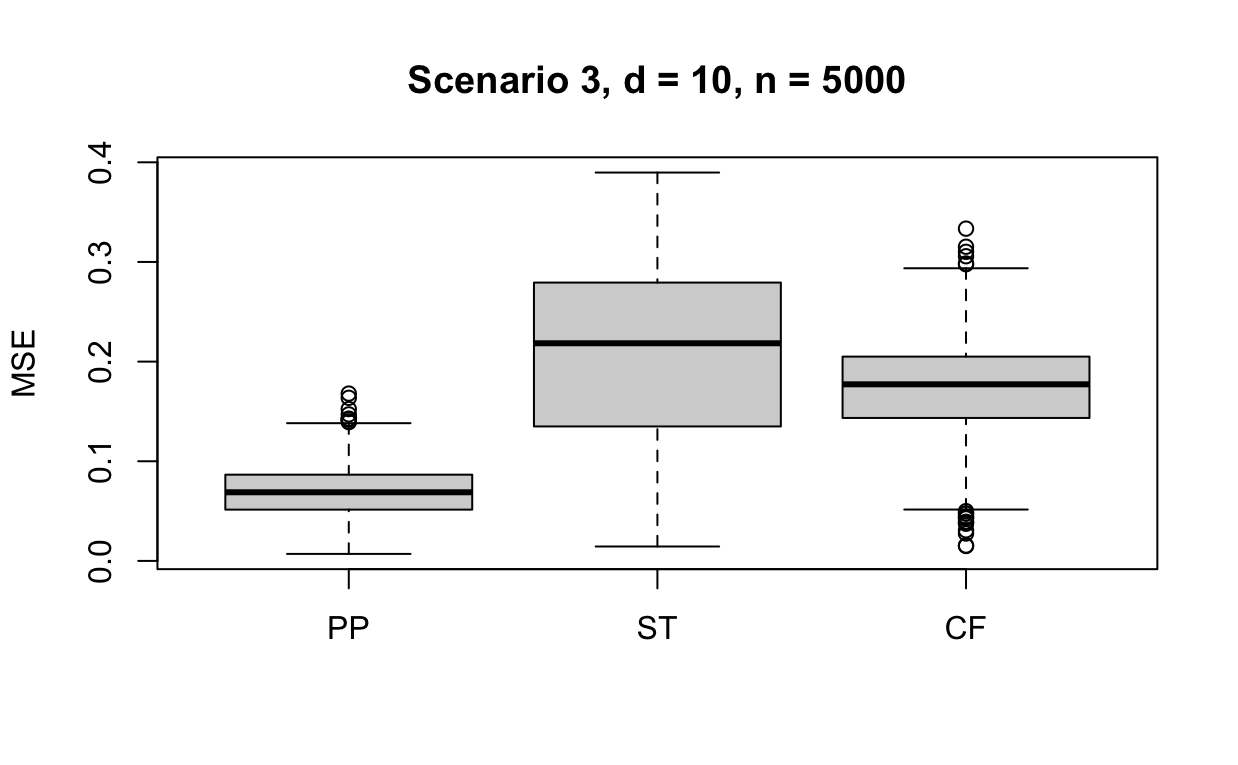}
\end{subfigure}\\
\begin{subfigure}{.45\textwidth}
  \centering
  \includegraphics[height = 36mm,width=.9\linewidth]{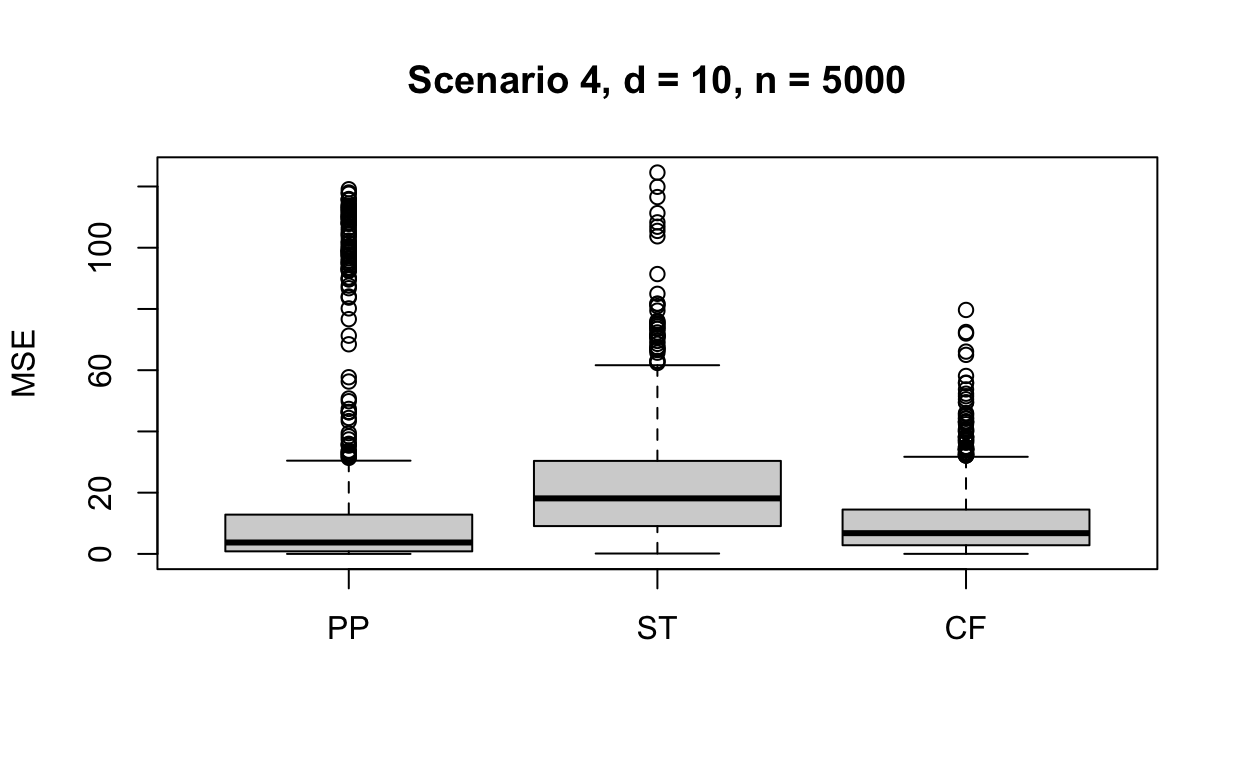}
\end{subfigure}%
\begin{subfigure}{.45\textwidth}
  \centering
  \includegraphics[height = 36mm,width=.9\linewidth]{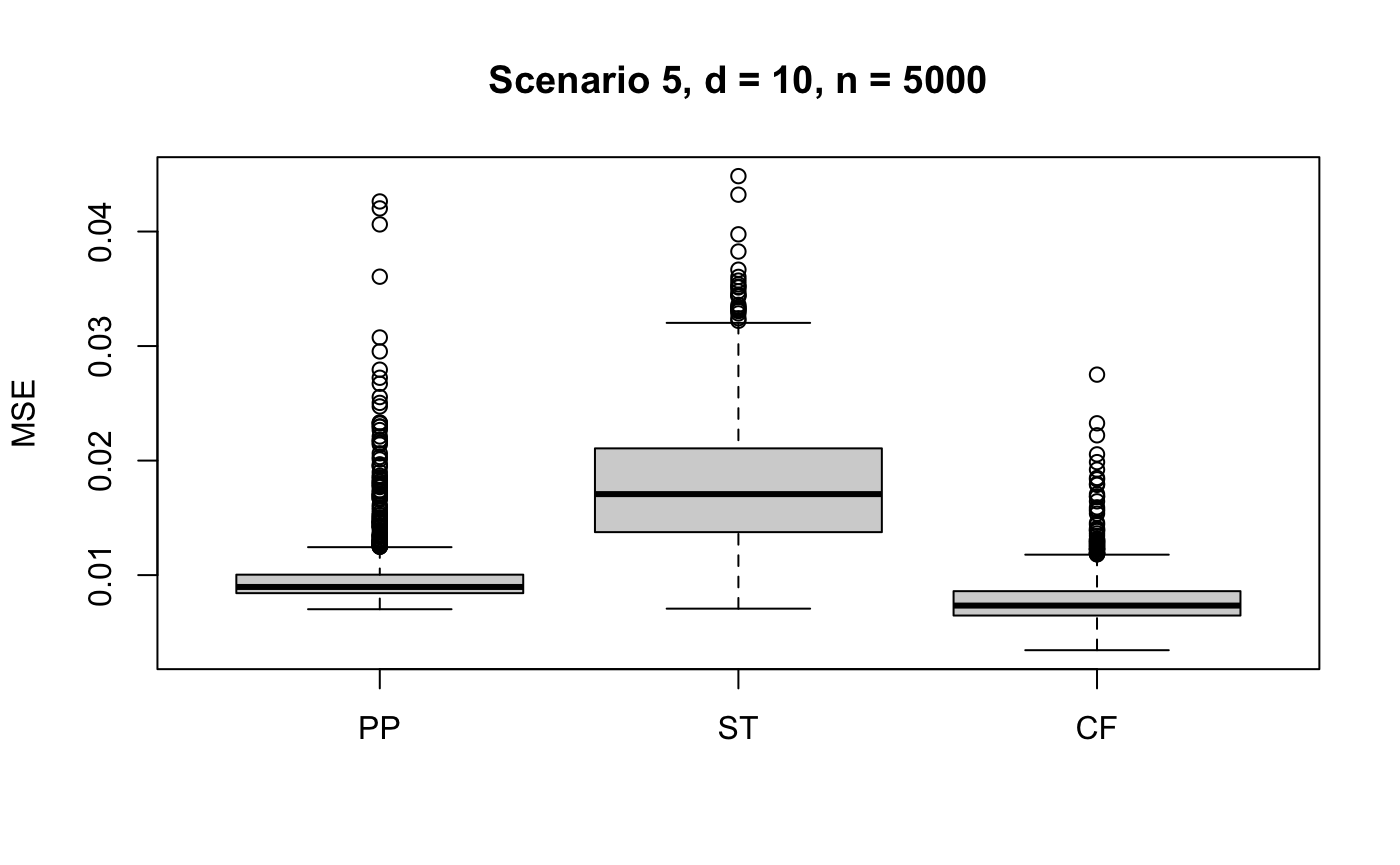}
\end{subfigure}
\\
\begin{subfigure}{.45\textwidth}
  \centering
  \includegraphics[height = 36mm, width=.9\linewidth]{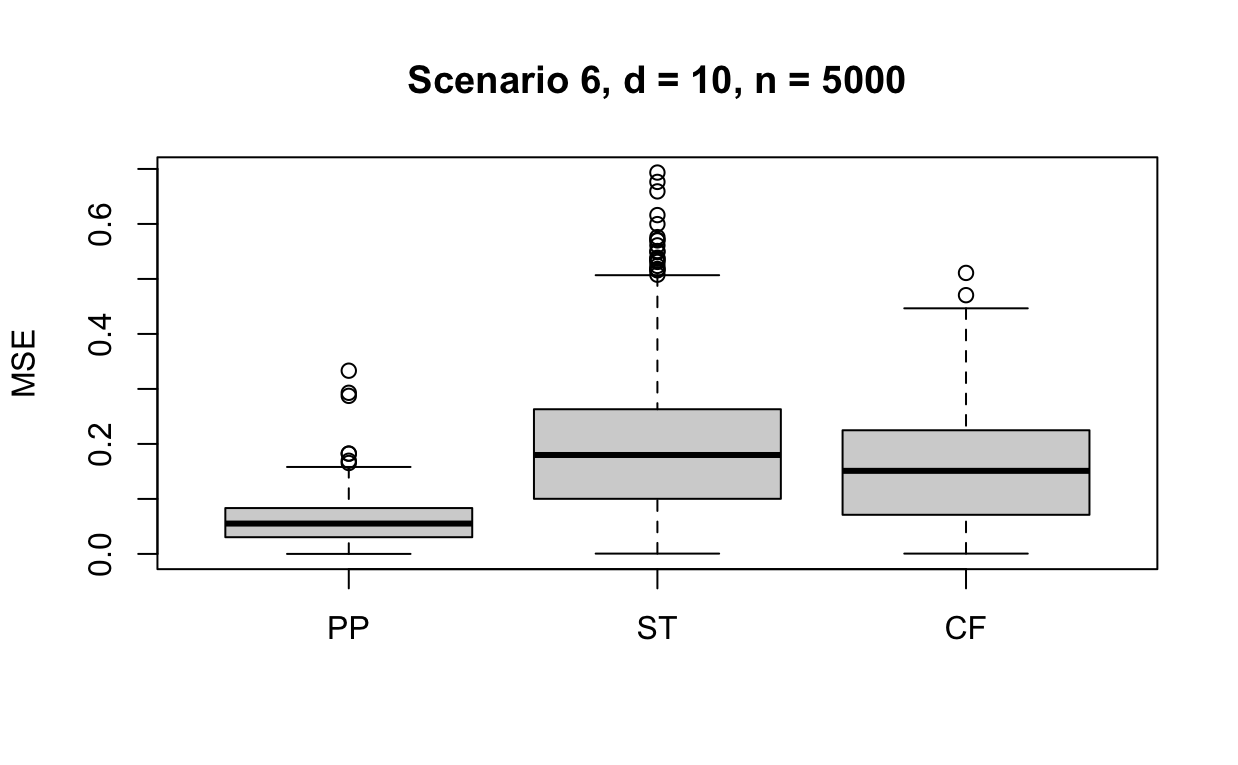}
\end{subfigure}%
\begin{subfigure}{.45\textwidth}
  \centering
  \includegraphics[height = 36mm,width=.9\linewidth]{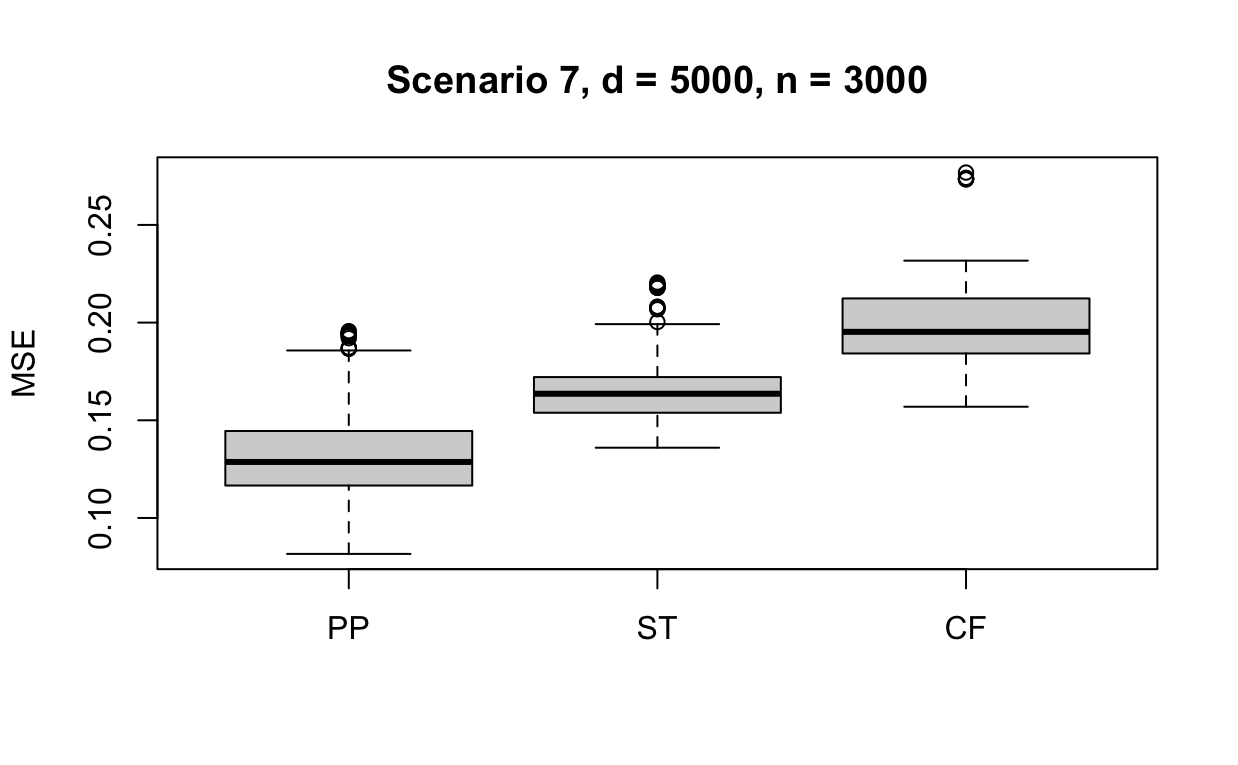}
\end{subfigure}
\captionsetup{justification=centering}
\caption{Comparison of mean squared errors over 1000 Monte Carlo simulations under different generative models for our proposed method (PP), endogenous stratification (ST) and causal forest (CF).}
\label{fig2}
\end{figure}

The boxplots that depict the distribution of MSEs obtained under all scenarios are presented in Figure \ref{fig2}. We can see that for Scenario 1, our proposed estimator achieves better accuracy when the sample size $n$ is large, and it is the best among the three estimators in these cases. The good performance of our method is consistent when we assume a more complex partition on the defined 2d grid as in Scenario 3. In addition, variation in accuracy, measured by the difference between the upper quartiles and the lower quartiles (also referred as the interquartile ranges) in each boxplot becomes smaller, accompanied with the increase in $n$. The reason for the enhanced performance when applying our method on a large-size data is due to the nature of nearest neighbor algorithm, which benefits more in learning the boundaries of clusters accurately with a larger sample size. Moreover, causal random forest suffers more in accuracy especially as $d$ gets larger \citep{Wager18}, while our method does not impact significantly because it focuses on a lower-dimensional space of the two scores instead of the overall covariate space. In Scenario 2, we introduce non-additivity and non-linear terms into the data model. Although linear assumptions are violated for both propensity and prognostic models, our method performs better compared to the other two methods regarding accuracy and variability. For a potential outcome model with randomized assignment of treatment and constant treatment effects, as in Scenario 4, our method has the best accuracy among all candidates, even though large noise is added to the true signal. In Scenario 5, our estimator performs slightly worse than causal forests in accuracy since the true treatment function is not directly dependent on propensity and prognostic scores, but it still outperforms endogenous stratification. However, for the previous two cases, our method shows larger variation in accuracy than the others. One of the reasons for this phenomenon is due to the misspecification in estimating the true scores and the large variance in constructing tree models. When the assumption of a sharp stratification of treatment effects in the 2d space of the two scores is invalid, as in Scenario 6, our method maintains its superiority over the benchmarks. With subgroups identified over the 2d interval of the scores, the mean squared errors are reduced through our estimator comparably better than the other methods. In a high-dimensional setting such as Scenario 7, we consider modified methods with lasso-regularized regressions for both our methodology and endogenous stratification, and our method remains its high performance as in the low-dimensional set-ups since the piecewise constant structure in treatment effects in the setting does not change. 

\begin{figure}[h!]
\centering
  \includegraphics[height = 41mm,width=.6\linewidth]{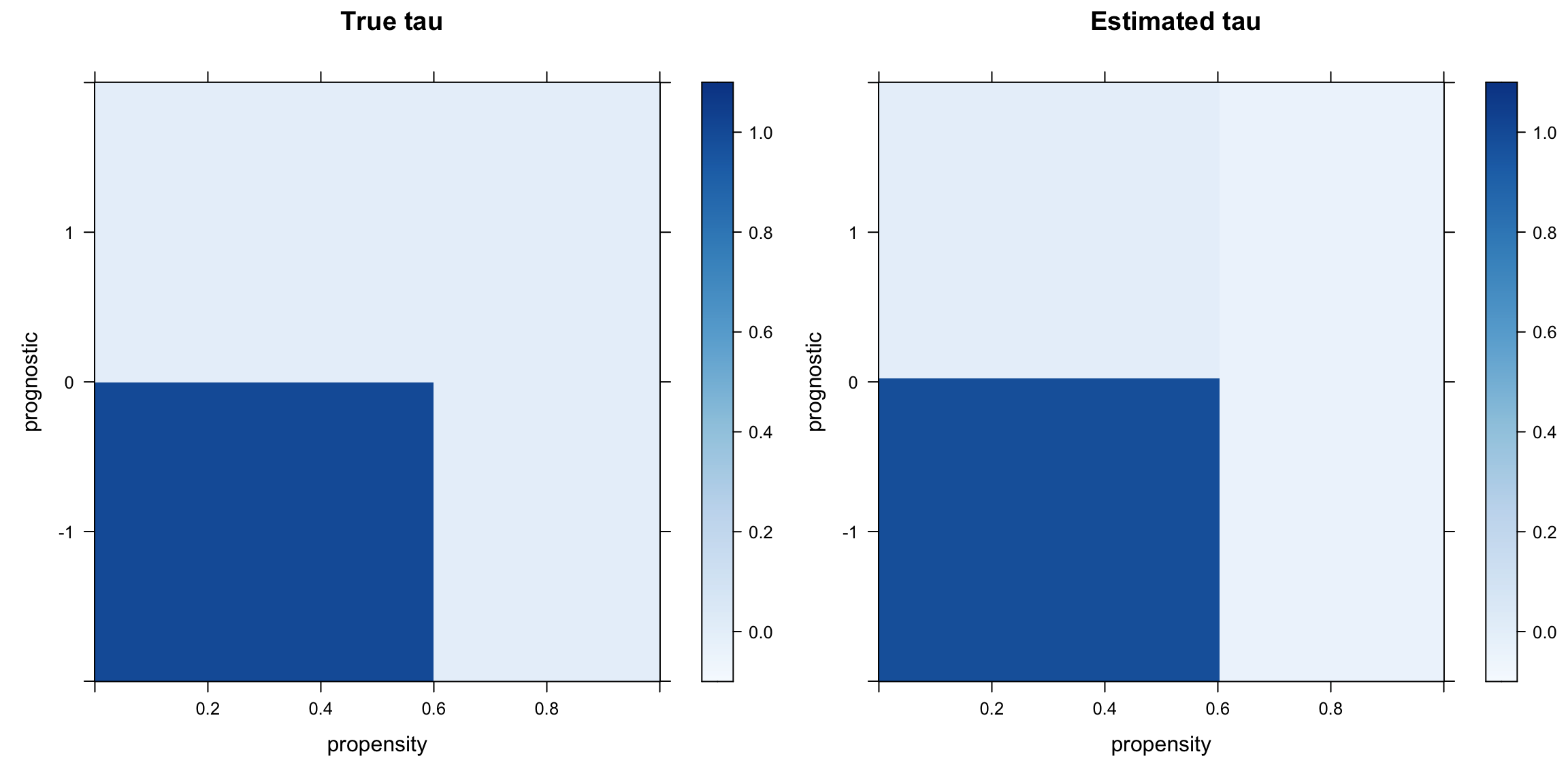}\\
    \includegraphics[height = 41mm,width=.6\linewidth]{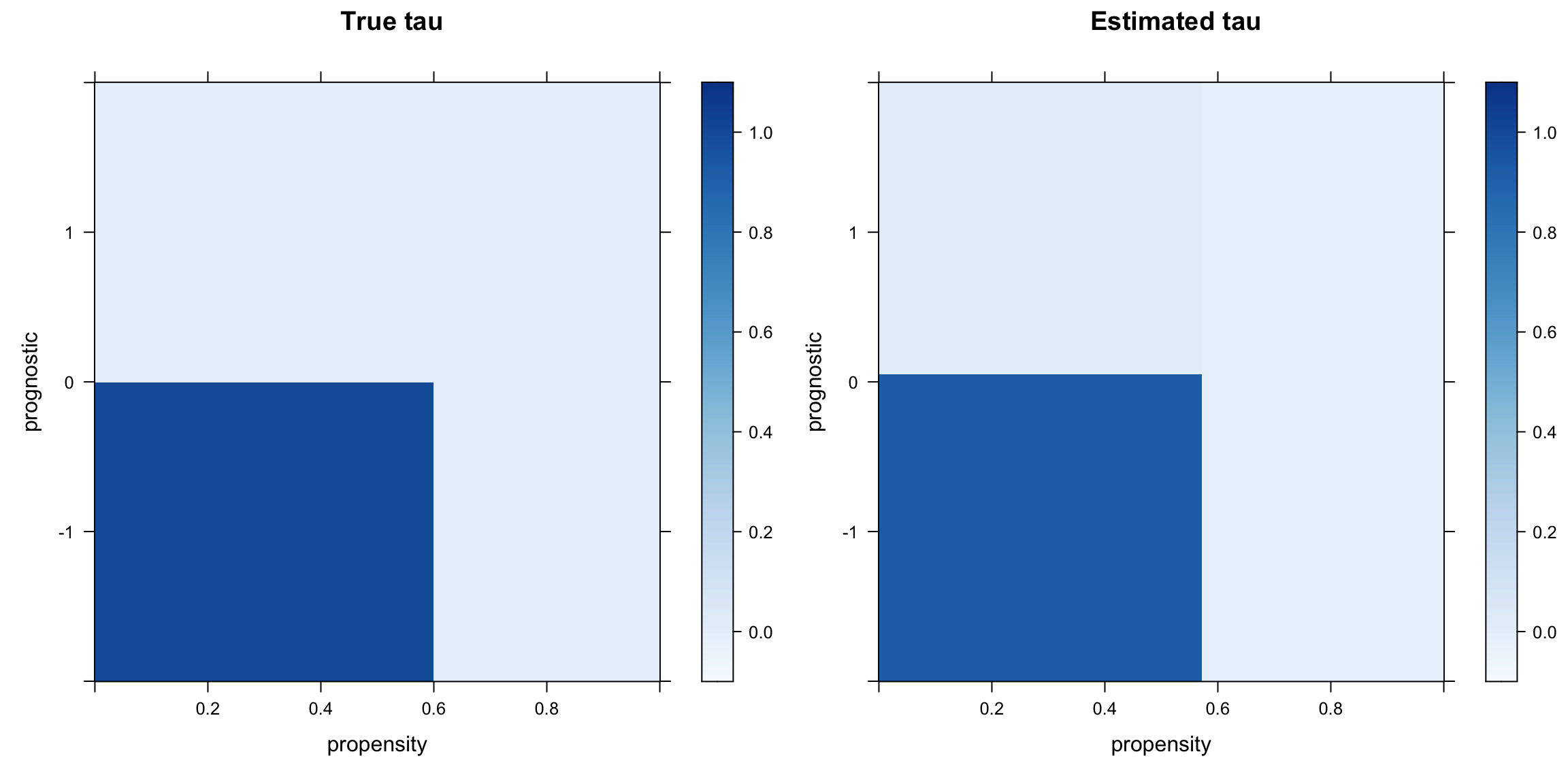}\\
    \includegraphics[height = 41mm,width=.6\linewidth]{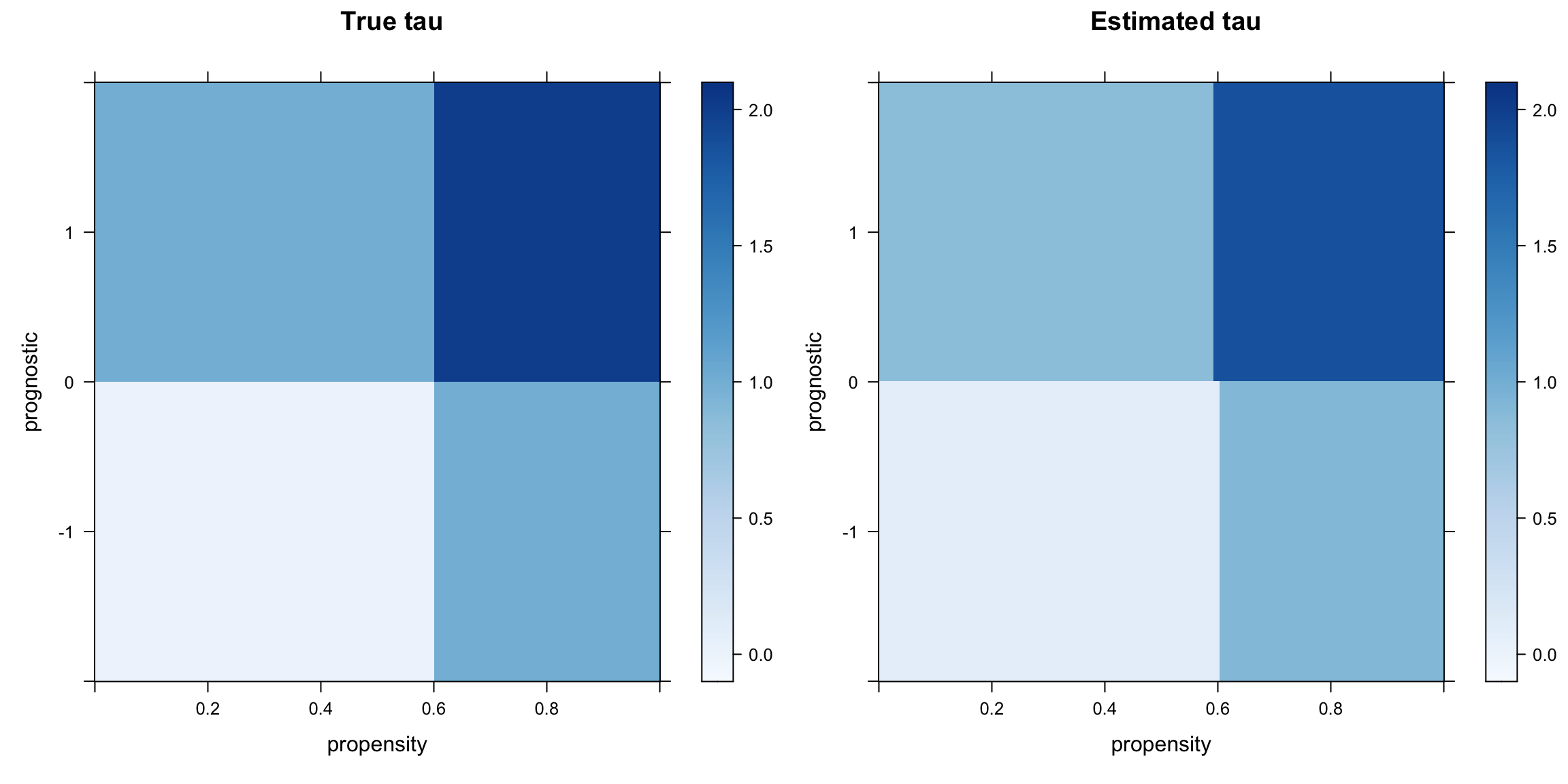}\\
        \includegraphics[height = 41mm,width=.6\linewidth]{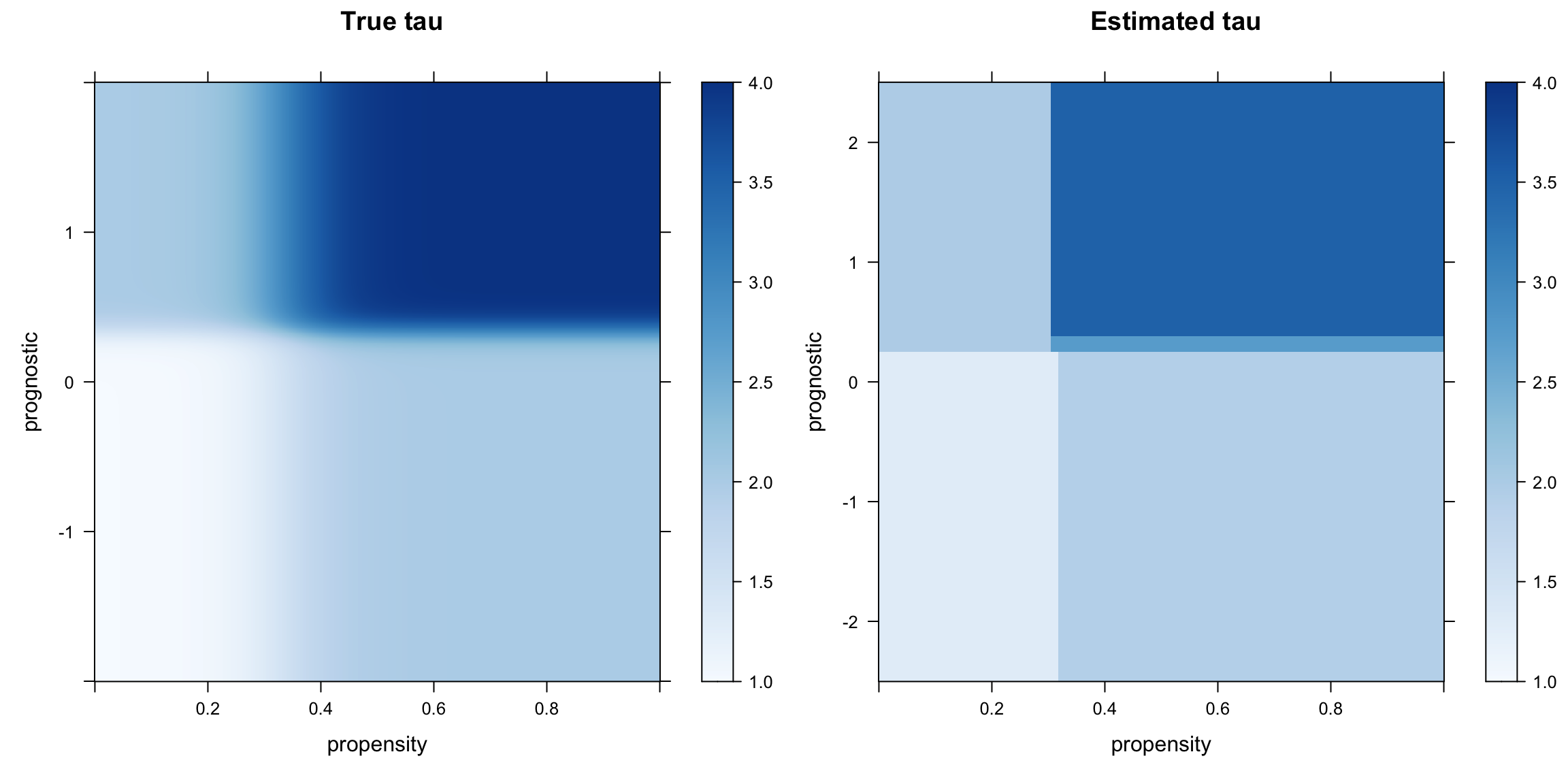}\\
     \includegraphics[height = 41mm,width=.6\linewidth]{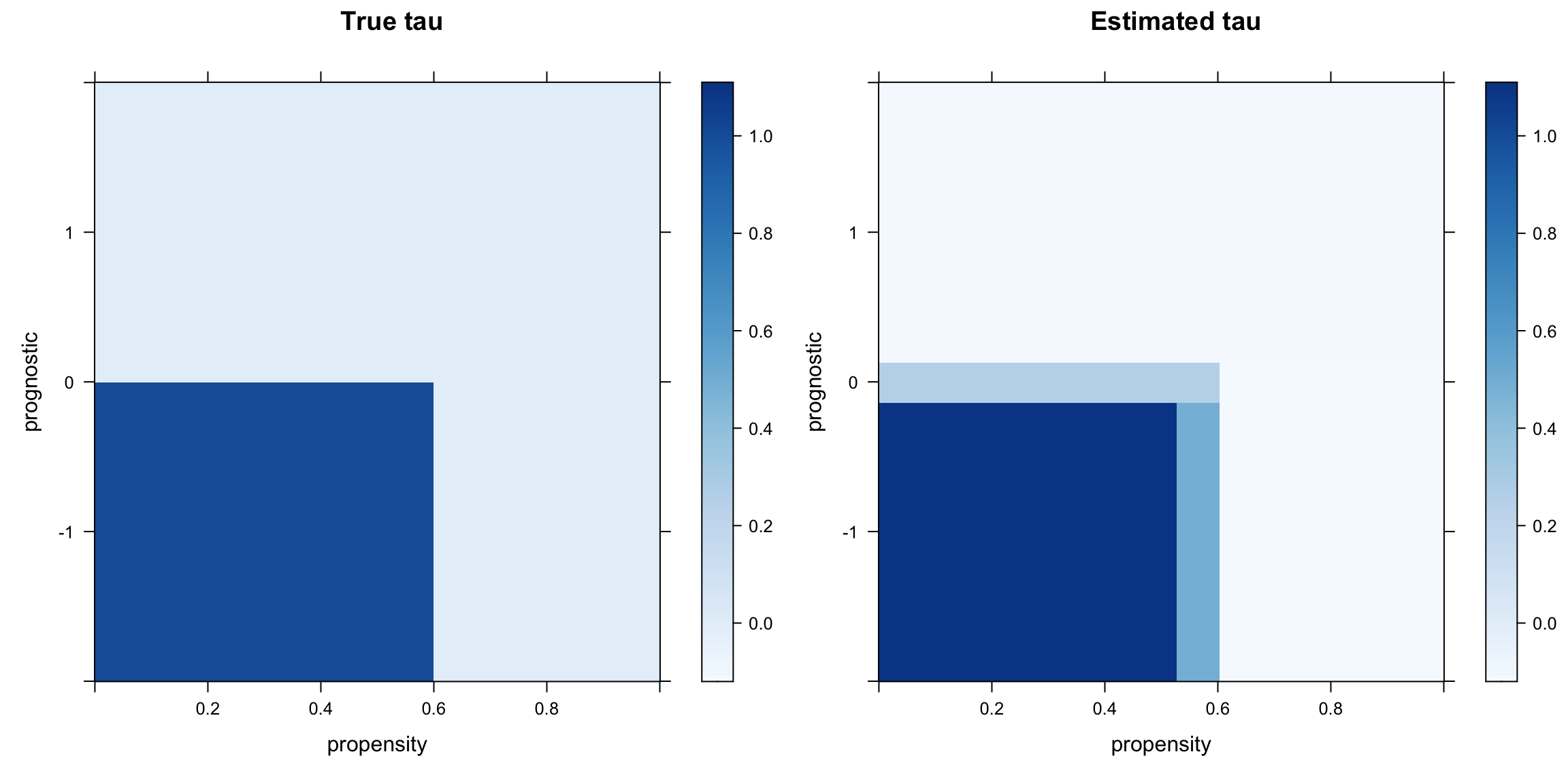}
\caption{One instance comparison between the true treatment effects and the estimates from the score-based method for Scenarios 1, 2, 3, 6 and 7 (from the top to the bottom). Plots on the left column depict the true treatment effect on the 2d grid of propensity and prognostic scores, and the right plots demonstrate the estimated treatment effects from our proposed method over the same grid.}
\label{fig3}
\end{figure}

We now take a careful look at the visualization comparison between the true treatment effect and the predictions obtained from our method for Scenarios 1, 2, 3, 6 and 7. We confine both the true signal and the predictive model in a 2d grid scaled on the true propensity and prognostic scores, as shown in Figure \ref{fig3}. It is not surprising that our proposed estimators provide a descent recovery of the piecewise constant partition in the true treatment effects over the 2d grid as in Scenarios 1, 2, 3, and 7, with only a small difference in the magnitude of treatment effects. In the setting when treatment effects are smoothly distributed across the grid of the two scores as in Scenario 6, our estimation does not exactly capture the soft boundary of the shape, but still provides us with a simplified delineation that help us understand general heterogeneity in treatment effects over the 2d space. 

With regard to uncertainty quantification, we examine coverage rates with a target confidence level of 0.95 for each method under different scenarios, and the corresponding results, along with the within iteration standard error estimates, are recorded in Table \ref{tab1}. It is quite clear that our proposed method achieves nominal coverage over the other two methods in almost all scenarios. It is worth to notice that although our estimator achieves a better coverage rate than the benchmarks, it suffers from large variation due to the nature of nearest-neighbor matching, especially in the case with small sample size. In general, our method that incorporates both propensity and prognostic scores in identifying subgroups and estimating treatment effects show its advantage in reducing bias, but the resulting large variance is worrisome and it remains space for future study to solve this issue.

\begin{table*}[ht!]
    \captionsetup{justification=centering}
    \caption{Reported coverage rates with a target confidence level of 0.95 and within iteration standard error estimates (in parentheses).}
    \centering
    \begin{tabular}{|c | c c | c c c |}
    \hline
        Scenario & $n$ & $d$ & PP & ST & CF \\
        \hline 
         1 & 1000 & 2 & 0.996 (0.177)  & 0.748 (0.158) & 0.969 (0.128) \\
          & 5000 & 10 & 0.987 (0.116) & 0.731 (0.062) & 0.736 (0.061) \\
         2 & 1000 & 10 & 0.994 (0.230) & 0.753 (0.152) & 0.797 (0.157) \\
         & 5000 & 10 & 0.956 (0.141) & 0.619 (0.064) & 0.829 (0.224) \\
         3 & 1000 & 10 & 0.989 (0.382) & 0.420 (0.163) & 0.769 (0.297) \\
         & 5000 & 10 & 0.949 (0.164) & 0.307 (0.075) & 0.750 (0.252)  \\
         4 & 1000 & 10 & 1.000 (28.621) & 0.933 (16.578) & 0.993 (18.067)\\
         & 5000 & 10 & 1.000 (6.693)& 0.998(6.049) & 0.973 (5.957)\\
         5 & 1000 & 10 & 1.000 (0.524) &  0.950 (0.353) & 0.978 (0.429) \\
         &  5000 & 10 & 0.845 (0.336) & 0.622 (0.287) & 0.716 (0.321)\\
         6 & 1000 & 10 & 0.995 (0.472) & 0.495 (0.161) & 0.865 (0.474)\\
         & 5000 & 10 & 0.999 (0.179)& 0.578 (0.089) & 0.821 (0.259)\\
         \hline
    \end{tabular}
    \label{tab1}
\end{table*}

We also compare the computational cost of our proposed scored-based method and causal forest, with simulated data from Scenario 1. Endogenous stratification is not considered in this comparison because the method exerts an arbitrary subclassification into a predetermined number of subgroups and it does not involve a cross-validation process. We record the averaged time consumed to obtain the estimate over 1000 simulations for both methods. Figure 4 demonstrates that our proposed method (PP) is comparably time-efficient over its competitor, and causal forest (CF) can be very expensive in operational time for large-size problems.

\begin{figure*}[ht!]
\centering
  \includegraphics[height = 75mm,width=.8\linewidth]{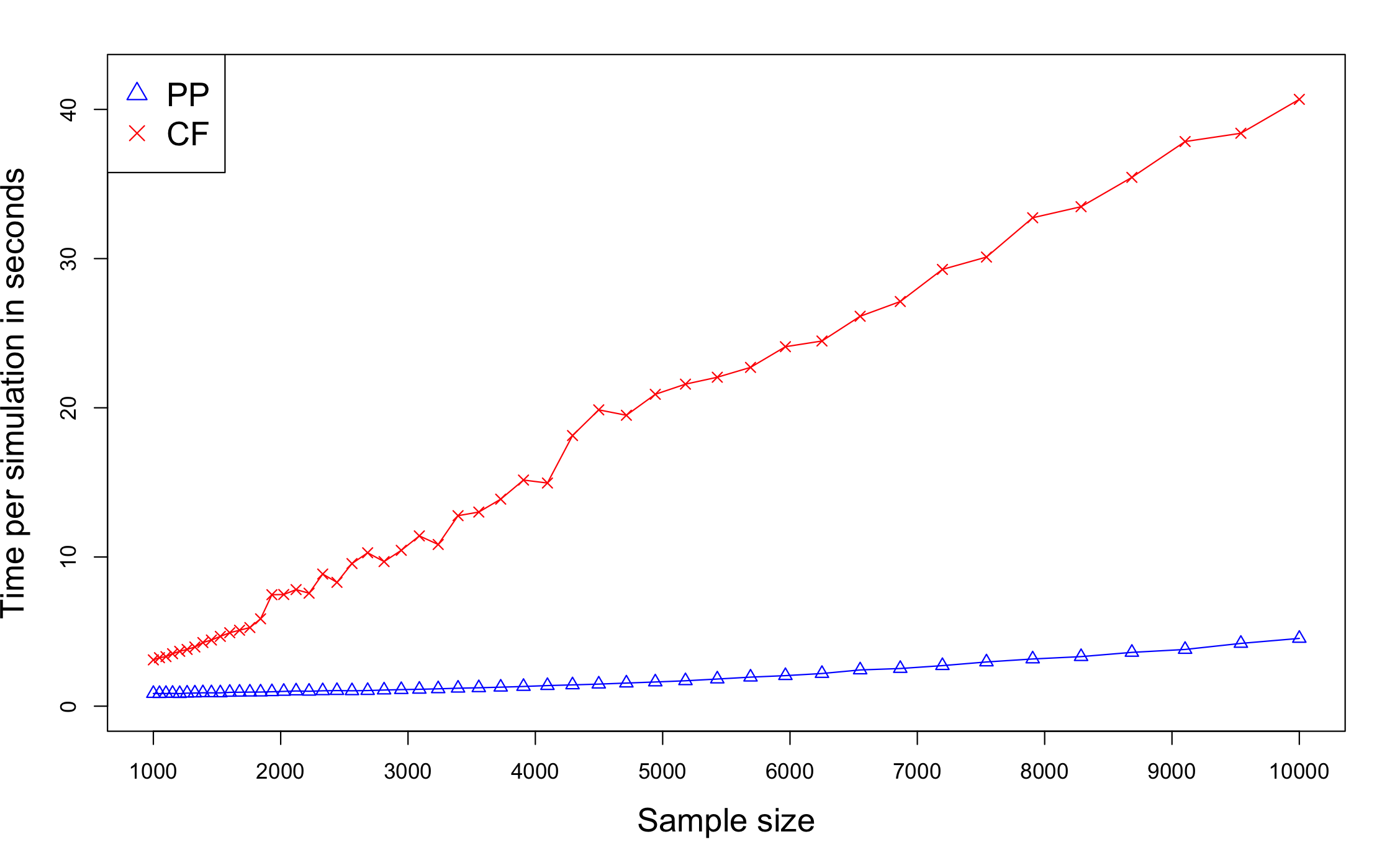}
  \captionsetup{justification=centering}
\caption{A plot of time per simulation in seconds for our proposed method (PP) and causal forest (CF) against problem size n (50 values from 1000 up to 10000). For each method, the time to compute the estimate for one simulated data is averaged over 1000 Monte Carlo simulations.}
\label{fig4}
\end{figure*}

These results highlight the promise of our method for accurate estimation of subgroups in heterogeneous treatment effects, all while emphasizing avenues for further work. An immediate challenge is to control the bias in estimation of propensity and prognostic scores. Using more powerful models instead of simple linear regression in estimation is a good way to reduce bias by enabling the trees to focus more closely on the coordinates with the greatest signal. The study of splitting rules for trees designed to estimate causal effects is still in its infancy and improvements may be possible.

\subsection{Real Data Analysis} 
To illustrate the behavior of our estimator, we apply our method on the two real-world data sets, one from a clinical study and the other from a complex social survey. Propensity score based methods are frequently used for confounding adjustment in observational studies, where baseline characteristics can affect the outcome of policy interventions. Therefore, the results from our method are expected to provide meaningful implications for these real data sets. However, one potential limitation of our proposed tree-based approach is that it imposes arbitrary dichotomization with sharp thresholds and thus may obscure potential continuous gradient over the space. For real data analysis in general, our method provides a reasonable simplification that is easy to interpret for the audience and gives some framework for thinking about when such a simplification would be appropriate.

\subsubsection{Right Heart Catheterization Analysis}
While randomized control trials (RCT) are widely encouraged as the ideal methodology for causal inference in clinical and medical research, the lack of randomized data due to high costs and potential high risks leads to the studies based on observational data. In this section, we are interested in examining the association between the use of right heart catheterization (RHC) during the first 24 hours of care in the intensive care unit (ICU) and the short-term survival conditions of the patients. Right Heart Catheterization (RHC) is a procedure for directly measuring how well the heart is pumping blood to the lungs. RHC is often applied to critically ill patients for directing immediate and subsequent treatment. However, RHC imposes a small risk of causing serious complications when administering the procedure. Therefore, the use of RHC is controversial among practitioners, and scientists want to statistically validate the causal effects of RHC treatments. The causal study using observational data can be dated back to \citet{Connors96}, where the authors implemented propensity score matching and concluded that RHC treatment lead to lower survival than not performing the treatment. Later, \citet{Hirano01} proposed a more efficient propensity-score based method and the recent study by \citet{Loh21} using a modified propensity score model suggested RHC significantly affected mortality rate in a short-term period. 

A dataset for analysis was first used in \citet{Connors96}, and it is suitable for the purpose of applying our method because of its its extremely well-balanced distribution of confounders across levels of the treatment \citep{Smith21}. The treatment variable $Z$ in the data indicates whether or not a patient received a RHC within 24 hours of admission. The binary outcome $Y$ is defined based on whether a patient died at any time up to 180 days since admission. The original data consisted of 5735 participants with 73 covariates. We preprocess the full data in the way suggested in \citet{Hirano01} and \citet{Loh21}, by removing all observations that contain null values in covariates,  dropping the singular covariate in the reduced data, and encoding categorical variables into dummy variables. The resulted data contains 2707 observations and 72 covariates, with 1103 in the treated group ($Z=1$) and 1604 in the control group ($Z=0$). Among the 72 observed covariates, there are 21 continuous, 25 binary, and 26 dummy variables transformed from the original 6 categorical variables. 

\begin{figure*}[ht!]
\centering
\begin{subfigure}{.6\textwidth}
  \centering
  \includegraphics[width=0.9\linewidth, height = 55mm]{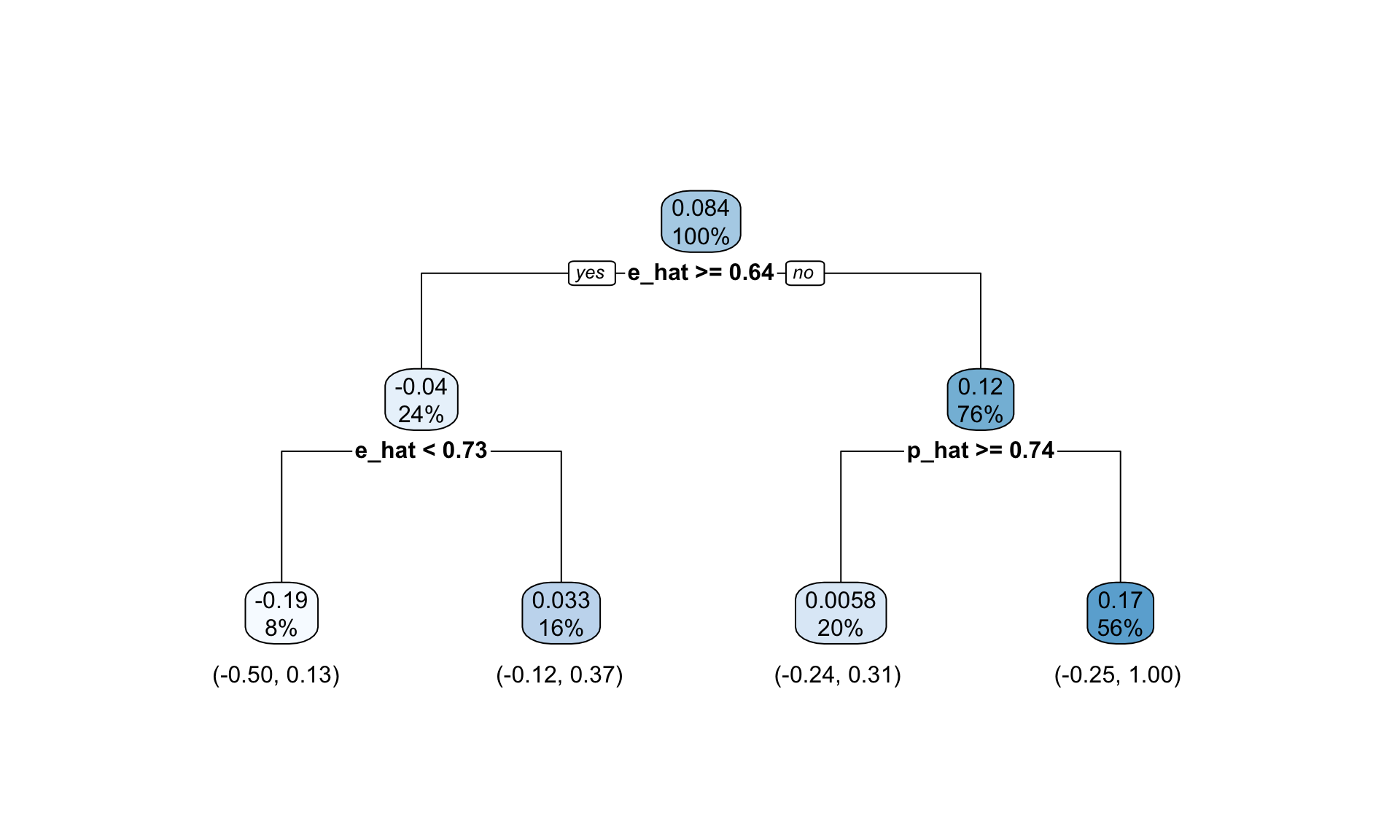}
\end{subfigure}%
\begin{subfigure}{.4\textwidth}
  \centering
  \includegraphics[width=.9\linewidth]{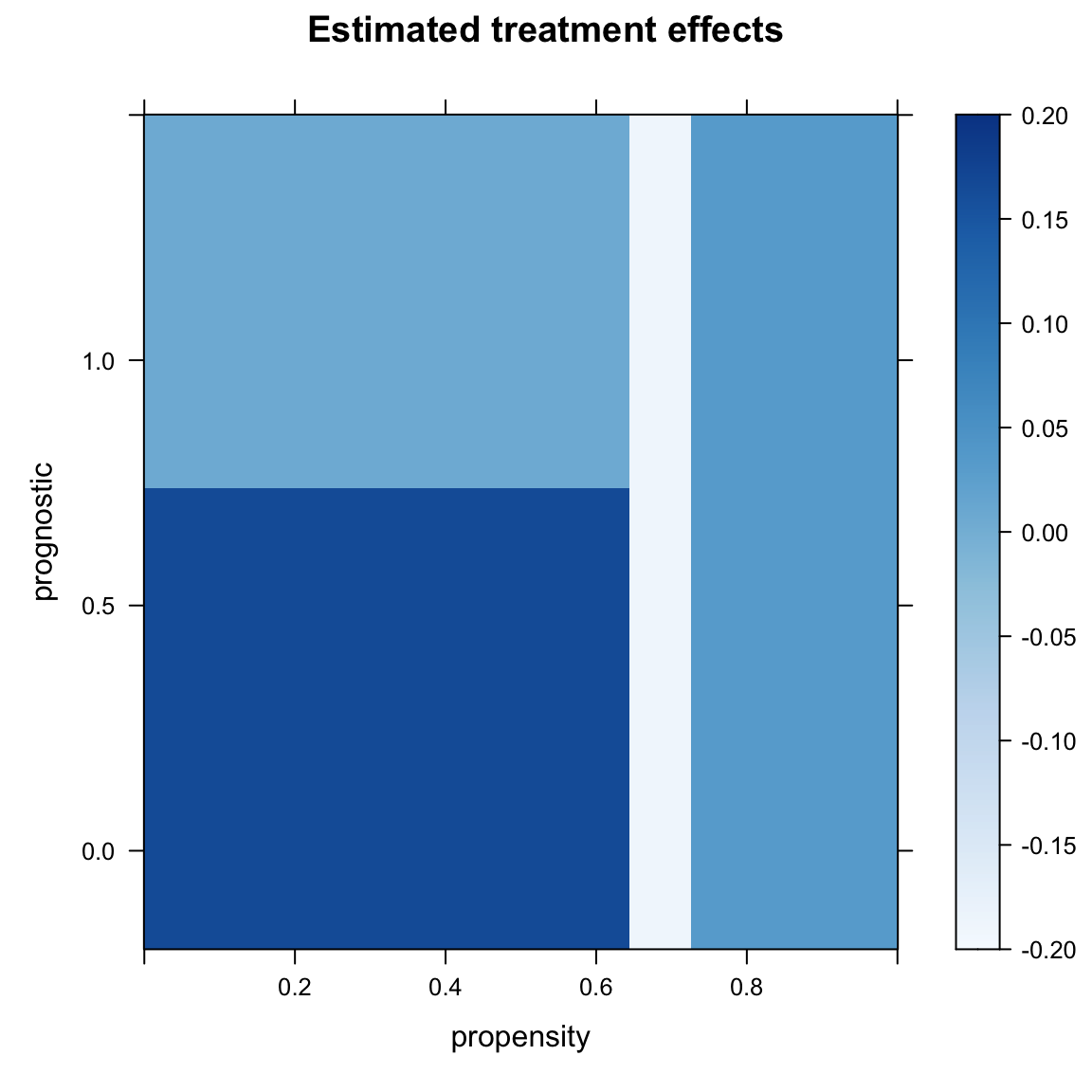}
\end{subfigure}
\captionsetup{justification=centering}
\caption{Prediction model for Right Heart Catheterization (RHC) data. \textit{Left:} a tree diagram of predictions on each split based on the two scores, with 95 $\%$ confidence intervals provided on the bottom. \textit{Right:} a 2d plot of prediction stratification on the intervals of propensity and prognostic scores, with a color scale on magnitude of treatment effects.}
\label{fig5}
\end{figure*}

The result of the prediction model from our proposed method is reported in Figure \ref{fig5} below. We observe that the sign of estimated treatment effects varies depending on the value of the propensity score and prognostic score. This particular pattern implies that RHC procedures indeed offer both benefits and risks in affecting patients' short-term survival conditions. Specifically, we are interested in the occurrence of large positive treatment effects (increase in chance of death) from the estimation. An estimated treatment effect of 0.17 (with a 95$\%$ confidence interval between -0.25 and 1.00) is observed on the group of patients with propensity scores less than 0.64 and prognostic scores less than 0.74, and this group accounts for 56$\%$ of the entire sample. For the rest of the sample, we observe relatively small treatment effects in magnitude. Under the scenario of RHC data, a smaller propensity score means that the patient is less likely to receive RHC procedures after admitting to the ICU, and it is related to the availability of RHC procedures at the hospital to which the patient is admitted. A smaller prognostic score tells that the patient has lower underlying chance of death. One possible explanation for this significant positive treatment on this certain group is that drastic change in treatment procedures that were applied to patients who do not actually need the aggressive style of care largely undermine patients' health conditions after admission and increase the mortality rate. In summary, our findings generally agree with the results and explanations in \citet{Connors96} and they offer some insights for practitioners to decide whether they should apply RHC procedures to patients.

\subsubsection{National Medical Expenditure Survey}
For the next experiment, we analyze a complex social survey data. In many complex surveys, data are not usually well-balanced due to potential biased sampling procedure. To incorporate score-based methods with complex survey data requires an appropriate estimation on propensity and prognostic scores. \citet{DuGoff14} suggested that combining a propensity score method and survey weighting is necessary to achieve unbiased treatment effect estimates that are generalizable to the original survey target population. \citet{Austin18} conducted numerical experiments and showed that greater balance in measured baseline covariates and decreased bias is observed when natural retained weights are used in propensity score matching. Therefore, we include sampling weight as an baseline covariate when estimating propensity and prognostic scores in our analysis.

In this study, we aim to answer the research question: how smoking habit affects medical expenditure over lifetime, and we use the same data set as in \citet{Johnson03}, which is originally extracted from the 1987 National Medical Expenditure Survey (NMES). The NMES included detailed information about frequency and duration of smoking with a large nationally representative data base of nearly 30,000 adults, and that 1987 medical costs are verified by multiple interviews and additional data from clinicians and hospitals. A large amount of literature focus on applying various statistical methods to analyze the causal effects of smoking on medical expenditures using the NMES data. In the original study by \citet{Johnson03}, the authors first estimated the effects of smoking on certain diseases and then examined how much those diseases increased medical costs. In contrast, \citet{Rubin01}, \citet{ Imai04} and \citet{Zhao20} proposed to directly estimate the effects of smoking on medical expenditures using propensity-score based matching and subclassification. \citet{Hahn20} applied Bayesian regression tree models to assess heterogeneous treatment effects.

For our analysis, we explore the effects of extensive exposure to cigarettes on medical expenditures, and we use pack-year as a measurement of cigarette measurement, the same as in \citet{Imai04} and \citet{Hahn20}. Pack-year is a clinical quantification of cigarette smoking used to measure a person's exposure to tobacco, defined by
$$\text{pack-year} = \frac{\text{number of cigarettes per day}}{20} \times \text{number of years smoked}.$$
Following that, we determine the treatment indicator $Z$ by the question whether the observation has a heavy lifetime smoking habit, which we define to be greater than 17 pack-years, the equivalent of 17 years of pack-a-day smoking.

The subject-level covariates $X$ in our analysis include age at the times of the survey (between 19 and 94), age when the individual started smoking, gender (male, female), race (white, black, other), marriage status (married, widowed, divorced, separated, never married), education level (college graduate, some college, high school graduate, other), census region (Northeast, Midwest, South, West), poverty status (poor, near poor, low income, middle income, high income), seat belt usage (rarely, sometimes, always/almost always), and sample weight. We select the natural logarithm of annual medical expenditures as the outcome variable $Y$ to maintain the assumption of heteroscedasticity in random errors. We preprocess the raw data set by omitting any observations with missing values in the covariates and excluding those who had zero medical expenditure. The resulting restricted data set contains 7903 individuals, with 4014 in the treated group ($Z=1$) and 3889 in the controlled group ($Z=0$). 

\begin{figure*}[ht!]
\centering
\begin{subfigure}{.6\textwidth}
  \centering
  \includegraphics[width=\linewidth]{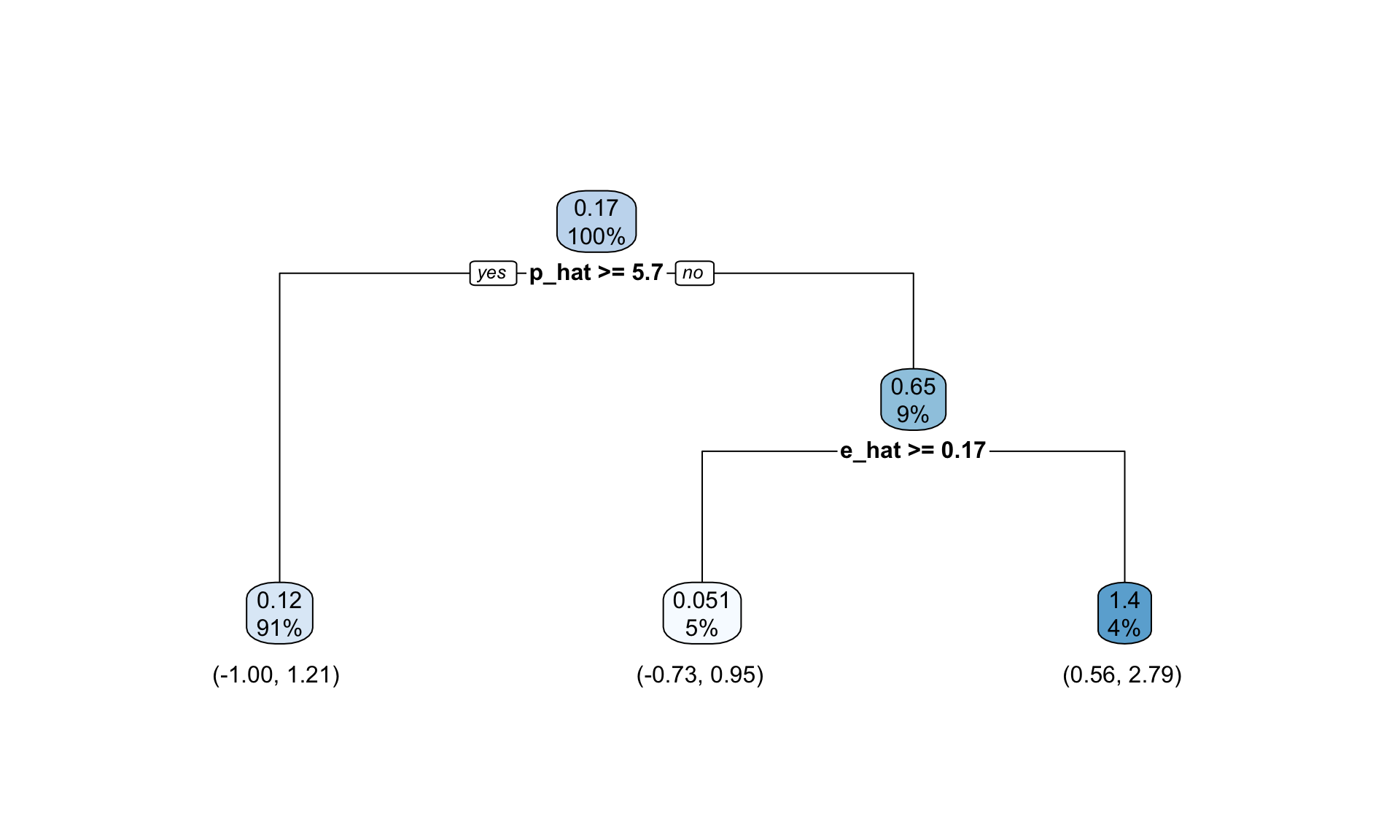}
\end{subfigure}%
\begin{subfigure}{.4\textwidth}
  \centering
  \includegraphics[width=.9\linewidth]{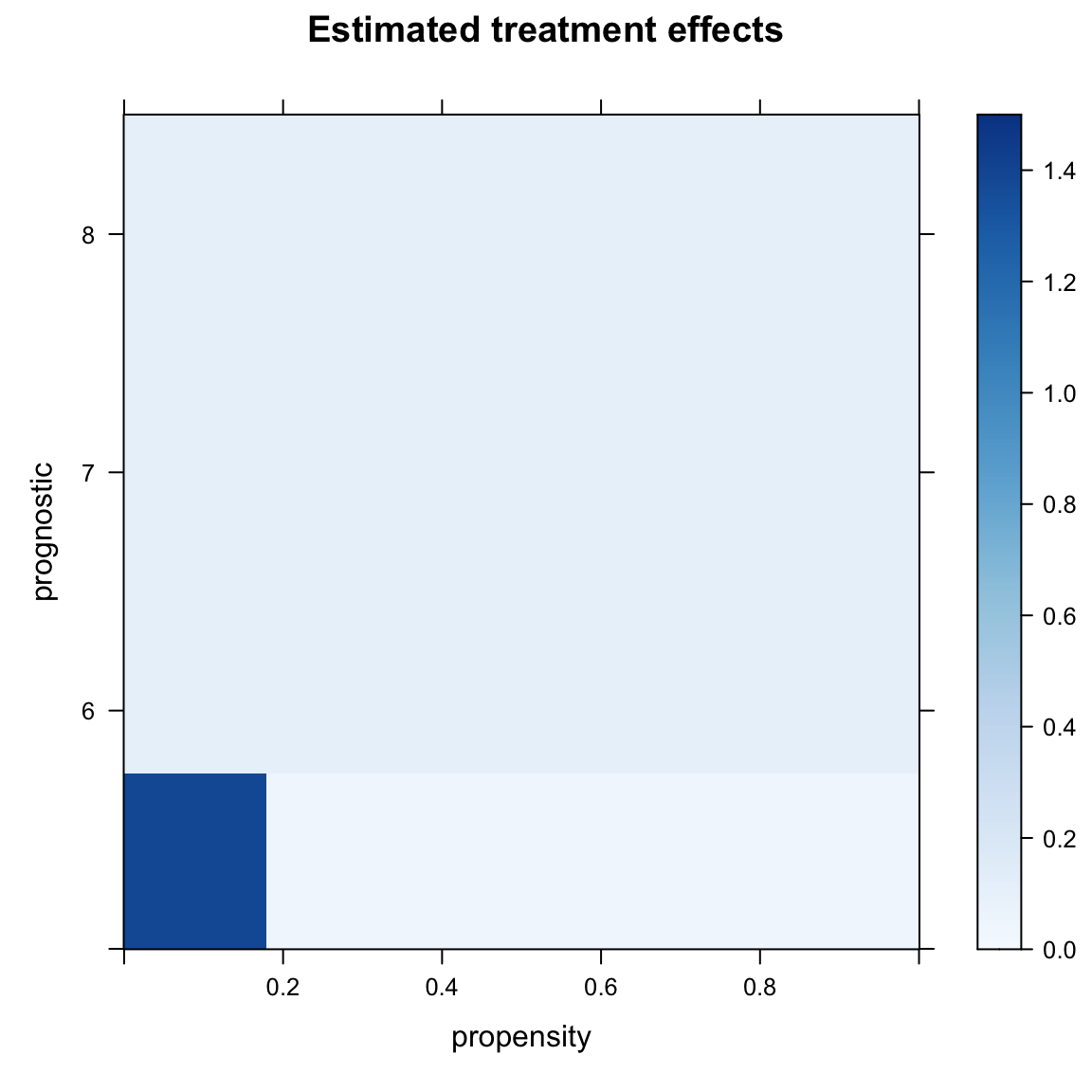}
\end{subfigure}
\captionsetup{justification=centering}
\caption{Prediction model for National Medical Expenditure Survey (NMES) data. \textit{Left:} a tree diagram of predictions on each split based on the two scores, with 95 $\%$ confidence intervals provided on the bottom. \textit{Right:} a 2d plot of prediction stratification on the intervals of propensity and prognostic scores, with a color scale on magnitude of treatment effects.}
\label{fig6}
\end{figure*}

The prediction model obtained from our method, as shown in Figure \ref{fig6}, is simple and easy to interpret. We derive a positive treatment effect across the entire sample, and the effect becomes significant when the predicted potential outcome is relatively low (less than 5.7). These results indicate that more reliance on smoking will deteriorate one's health condition, especially for those who currently do not have a large amount of medical expenditure. Moreover, we observe a significant positive treatment effect of 1.4 (with a 95$\%$ confidence interval between 0.56 and 2.79), in other word, a certain and substantial increase in medical expenditure for the subgroup with propensity score less than 0.17. It is intuitive to assume that a smaller possibility of engaging in excessive tobacco exposure is associated with healthier living styles. This phenomenon is another evidence that individuals who are more likely to stay healthy may suffer more from excessive exposure to tobacco products. In all, these results support policymakers and social activists who advocate for nationwide smoking ban.

\section{Discussions}
\label{sec6}
Our method is different from existing methods on estimating heterogeneous treatment effects in a way that we incorporate both matching algorithms and non-parametric regression trees in estimation, and the final estimate can be regarded as a 2d summary on treatment effects. Moreover, our method exercises a simultaneous stratification across the entire population into subgroups with the same treatment effects. Subgroup treatment effect analysis is an important but challenging research topic in observational studies, and our method can be served as an efficient tool to reach a reasonable partition. 

Our numerical experiments on various simulated and real-life data lay out empirical evidence of the superiority of our estimator over state-of-the-art methods in accuracy and our proposal provides insightful interpretation on heterogeneity in treatment effects across different subgroups. We also discovered that our method is powerful in investigating subpopulations with significant treatment effects. Identifying representative subpopulations that receive extreme results after treatment is a paramount task in many practical contexts. Through empirical experiments on two real-world data sets from observational studies, our method demonstrates its ability in identifying these significant effects.

Although our method shows its outstanding performance in estimating treatments effects under the piecewise constant structure assumption, it remains meaningful and requires further study to develop more accurate recovery of such structure. For example, a potential shortcoming of using conventional regression trees for subclassification is that the binary partition over the true signals is not necessarily unique. Using some variants of CART, like optimal trees \citep{Bertsimas17} and dyadic regression trees \citep{Donoho97}, would be more appropriate for estimation under additional assumptions. Another particular concern with tree-based methods is the curse of dimensionality. Applying other non-parametric regression techniques, such as $K$-nearest-neighbor fused lasso \citep{Padilla20} and locally estimated scatterplot smoothing (LOESS), may offer a compromised solution to deal with this issue and learn a more smoothed structure in treatment effects other than a rectangular partition on 2d data. It is also worth improving the estimation of propensity and prognostic scores using similar non-parametric based methods if a piecewise constant assumption hold for the two scores as well. 

Moreover, we acknowledge that fitting propensity score and prognostic scores normally would demand more than main terms linear and logistic regressions as used in the current method to have hope of eliminating bias in estimating the two scores. The key takeaway of our proposal is not to estimate the two scores more accurately, but to provide a lower-dimensional summary on heterogeneous treatment effects. If more foreknowledge on the structure of the underlying functions of the two scores can be incorporated, we can certainly select a more ideal algorithm or machine learning model for estimation. To compromise the lack of such information under a usual scenario, a SuperLearner ensemble estimator, proposed by \citet{van07}, may offer a more promising solution for estimating the two scores by taking a weighted average of multiple prediction models. Yet, an important challenge for future work is to design rules that can automatically choose the best model to fit the data.

Another question to consider is the source of heterogeneity in causal effects. Our proposal that utilizes two scores can be seen as a reasonable starting point to analyze heterogeneity from a low-dimensional perspective, but it cannot comprehensively capture the sources of such effects. For instance, it is seemingly possible to exist a covariate that leads to substantial heterogeneous treatment effects on the response but does not itself have a large effect on propensity or prognostic scores. There is no particular guiding theories about this issue in current literature, and thus it remains huge space for further investigation.

\begin{appendix}
\section{Study on the choice of the number of nearest neighbors}
\label{appA}
In this section, we examine how the number of nearest neighbors in the matching algorithm affects the estimation accuracy. Recall in Step 2 of our proposed method, we implement a $K$-nearest-neighbor algorithm based on the two estimated scores for a sample of size $n$. The computational complexity of this $K$-NN algorithm is of $O(Kn)$. Although a larger $K$ produce a promisingly higher estimation accuracy, more computational costs become the corresponding side-effect. Moreover, the issue of ``underfitting'' - the increase of bias occurs when we make such selection. It is because a selection of too large $K$ defeats the underlying principle behind nearest neighbor algorithms - that neighbours that are nearer share similar densities. Therefore, a smart choice of $K$ is essential to balance the trade-off among accuracy, computational expense, and generability.

We follow the same generative model in Scenario 1 from Section \ref{sec5} and compute the averaged mean squared error over 1000 Monte Carlo simulations for $K = 1,...,50$ with a fixed sample size $n = 5000$. The results in Figure \ref{fig6} show that the averaged MSE continuously decreases as the number of nearest neighbors $K$ selected in the matching algorithm grows. However, the speed of improvement in accuracy gradually slows down when $K$ exceeds 10, which is close to $\log(5000)$. This suggests that an empirical choice of $K\approx \log(n)$ is sufficient to produce a reasonable estimate on the target parameter and this choice is more 'sensible' than the conventional setting of $K = 1$.

\begin{figure}[ht!]
  \centering
  \includegraphics[height = 65mm, width=.7\linewidth]{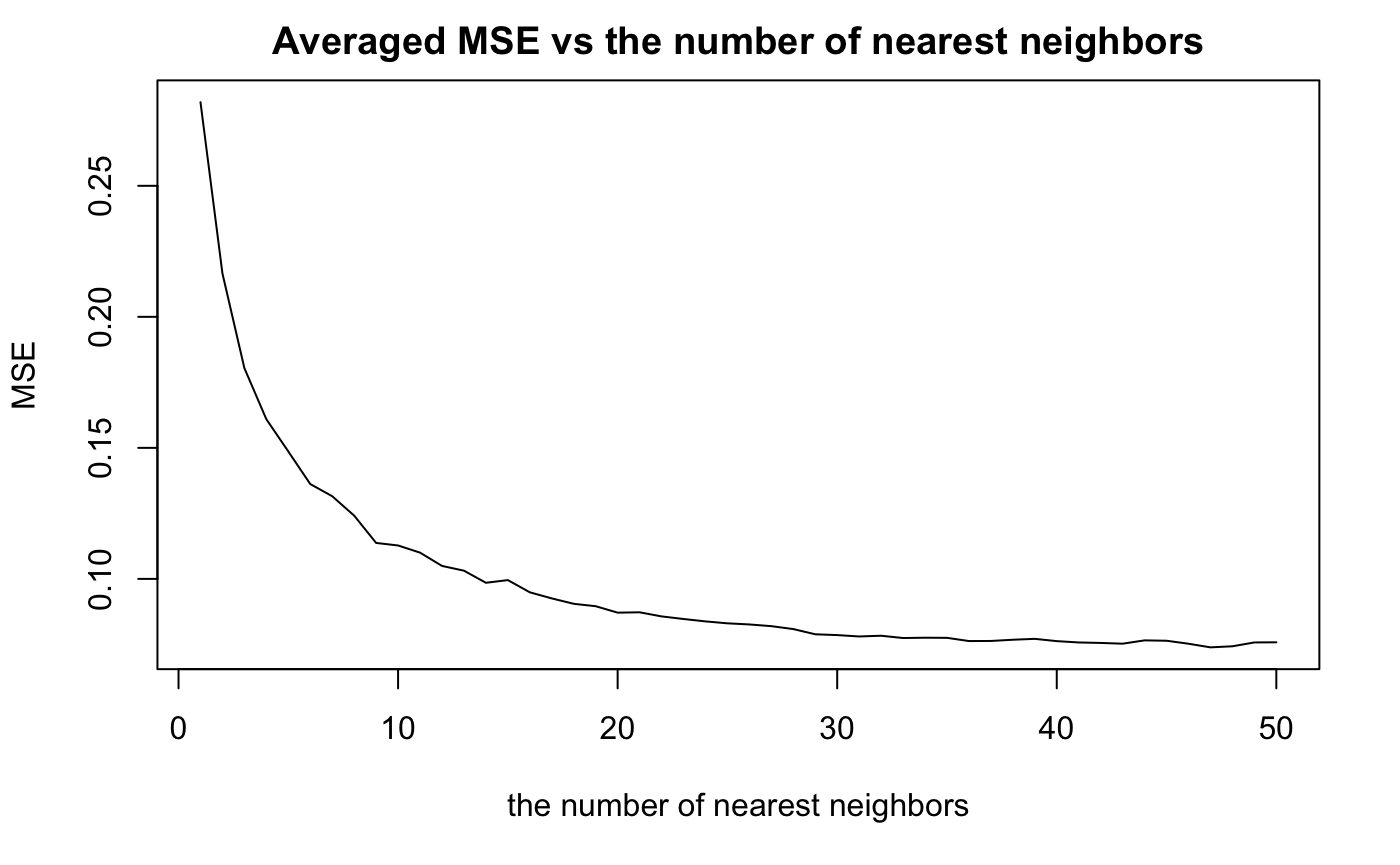}
  \captionsetup{justification=centering}
\caption{The plot of averaged MSE against the number of nearest neighbors.}
\label{fig7}
\end{figure}

\section{A Summary Table of Simulation Studies}
\label{appB}
In this appendix, we include a table to provide a text summary of all simulation models considered in Section 5.1. We add some notes to help the audience understand similarities, differences, and sources of these data generative models.
\begin{table*}[ht!]

    \centering
    \begin{tabular}{lp{3.5cm}p{3.5cm}p{3.6cm}p{3.5cm}}
 \textbf{Scenario} & \textbf{Propensity Model} & \textbf{Prognostic Model} & \textbf{Treatment Effect Model} & \textbf{Note} \\
 \hline
 1 & linearity in covariates & linearity in covariates & a sharp delineation over the 2d grid of propensity and prognostic scores with 2 subgroups & \\
 \hline
 2 & linear model with additional interactions and non-linear terms & linear model with additional interactions and  non-linear terms & a sharp delineation over the 2d grid of propensity and prognostic scoreswith 2 subgroups & the same treatment effect model as in Scenario 1; both propensity and prognostic scores are misspecified with linear model \\
 \hline
 3 & linearity in covariates & linearity in covariates  & a sharp delineation over the 2d grid of propensity and prognostic scores with 4 subgroups  & the same propensity and prognostic models as in Scenario 1; more subgroups in treatment effects   \\
 \hline
 4 & random with a fixed number of treatments &linearity in covariates with an intercept & constant zero & the same setting from \citet{Abadie18}\\
 \hline
 5 & a smooth function of single covariate & proportional to single covariate & a function of the covariates & modified from Equation 27 of \citet{Wager18}\\
 \hline
 6 & linearity in covariates & linearity in covariates & a smooth function of the two scores& modified from Equation 28 of \citet{Wager18}\\
  \hline
 7 & linearity in a few covariates with sparsity in others & linearity in a few covariates with sparsity in others & a sharp delineation over the 2d grid of propensity and prognostic scores with 2 subgroups  & a high-dimensional setting corresponding to Scenario 1\\
\end{tabular}
\end{table*}

\section{Non-parametric Bootstrap in Simulation Studies}
\label{appC}
In Section \ref{sec5}, we use non-parametric bootstrap to construct confidence intervals for endogenous stratification and our proposed method. We use these bootstrap samples to compute coverage rates with respect to a target level of $95\%$ as a measurement of uncertainty. The bootstrap method, introduced by \citep{Efron79}, is a simple but powerful tool to obtain a robust non-parametric estimate of the confidence intervals through sampling from the empirical distribution function of the observed data. It is worth noting that causal forests or generalized random forests can compute confidence intervals quickly without requiring a bootstrap approach on which our method relies.

In this appendix, we introduce the details on how we implement non-parametric bootstrap for the purpose of computing coverage rates in the simulation experiments. For each scenario in Section \ref{sec5},  we start with generating a sample following the defined data generation model with a sample size $n$. Next, we create 1000 random samples with replacement from this single set of data, also with the sample size $n$. We then apply both methods on these simulation repetitions, and obtain a series of estimations on each unit in the original set. Following these estimations, we calculate the corresponding $2.5\%$ and $97.5\%$ quantiles for all units in the original sample. Coverage rates of a $95\%$ prediction level are thus the frequencies of the original units falling inside the intervals between the two quantiles computed in the previous step.
\end{appendix}

\printbibliography
\end{document}